\documentclass[empty]{jfm}
\usepackage{graphbox,graphicx}
\usepackage{newtxtext}
\usepackage{newtxmath}
\usepackage{amsmath}
\usepackage{amssymb}
\usepackage{amsfonts}
\usepackage{natbib}
\usepackage{stmaryrd}
\usepackage{subcaption}
\usepackage{float}
\usepackage{bm}
\usepackage{soul}
\usepackage{hyperref}
\usepackage{orcidlink}
\hypersetup{
    colorlinks = true,
    urlcolor   = blue,
    citecolor  = blue,
}
\usepackage{color}
\usepackage{tikz}

\newcommand{\RomanNumeralCaps}[1]

\newcommand{\unitz}{{\boldsymbol e}_z}
\newcommand{\unity}{{\boldsymbol e}_y}
\newcommand{\unitx}{{\boldsymbol e}_x}

\newcommand{\unitr}{{\boldsymbol e}_r}
\newcommand{\unitt}{{\boldsymbol e}_\theta}

\newcommand{\up}{\boldsymbol{u}_\perp}

\newcommand{\vel}{\boldsymbol u}
\newcommand{\Dt}{D\!t}
\newcommand{\Dz}{D\!z}

\newcommand{\Ha}{H\!a}
\newcommand{\Rm}{R\!m}
\newcommand{\Lu}{L\!u}
\newcommand{\Rn}{R\!_\eta}
\newcommand{\Rv}{R\!_\nu}
\newcommand{\NOmega}{N_\omega}
\newcommand{\Pm}{P\!m}
\renewcommand{\Rey}{R\!e}
\newcommand{\Ja}{J\!a}
\newcommand{\Al}{A\!l}

\newcommand{\rmb}{\widetilde{\boldsymbol b}}
\newcommand{\rmbt}{\widetilde{b}_\theta}

\newcommand{\ie}{\emph{i.e.} }
\renewcommand{\eg}{\emph{e.g.} }
\renewcommand{\figurename}{Figure }

\newcommand{\mathbfitsf}[1]{\textit{\textsf{\textbf{#1}}}}

\title{Alfv\'en waves at low magnetic Reynolds number: Transitions between diffusion, dispersive Alfv\'en waves and nonlinear propagation}

\author{Samy Lalloz\aff{1},
Laurent Davoust\aff{2},
Fran\c cois Debray \aff{3} \and Alban Poth\'erat \aff{1} \corresp{\email{alban.potherat@coventry.ac.uk}}
}

\affiliation{\aff{1}  Coventry University, Centre for Fluid and Complex Systems, Priory Street, Coventry CV1 5FB, UK
\aff{2} SIMaP, Electromagnetic Processing of Materials (EPM) Laboratory, Grenoble-INP/CNRS/Université Grenoble-Alpes, F-38000 Grenoble, France
\aff{3} Laboratoire National des Champs Magn\'etiques Intenses (LNCMI), CNRS UPR 3228    , EMFL, Universit\'e Toulouse III - Paul Sabatier, Universit\'e F\'ed\'erale Toulouse Midi    -Pyr\'en\'ees, Institut National des Sciences Appliqu\'ees, Universit\'e Grenoble Alpes, 3    8042 Grenoble CEDEX, France}

\makeatletter
\newcommand{\vast}{\bBigg@{3.5}}
\newcommand{\Vast}{\bBigg@{4}}
\makeatother

\newcommand{\thickLine}{\raisebox{2pt}{\tikz{\draw[line width = 1.5pt] (0,0) -- (5mm,0);}}}
\newcommand{\thickDottedLine}{\raisebox{2pt}{\tikz{\draw[line width = 1.5pt, dotted] (0,0) -- (5mm,0);}}}
\newcommand{\normalLine}{\raisebox{2pt}{\tikz{\draw[line width = 1pt] (0,0) -- (5mm,0);}}}
\newcommand{\dashedLine}{\raisebox{2pt}{\tikz{\draw[dashed, line width = 1.0pt] (0,0) -- (5 mm,0);}}}
\newcommand{\dottedLine}{\raisebox{2pt}{\tikz{\draw[line width = 1.0pt, dotted] (0,0) -- (5 mm,0);}}}
\newcommand{\dashDottedLine}{\raisebox{2pt}{\tikz{\draw[line width = 1.0pt, dash dot] (0,0) -- (5 mm,0);}}}

\begin{document}
\maketitle

\begin{abstract}
	We seek the conditions in which Alfvén waves (AW) can be produced in laboratory-scale liquid metal experiments, i.e. at low magnetic Reynolds Number ($\Rm$). Alfvén waves are incompressible waves propagating along magnetic fields typically found in geophysical and astrophysical systems. Despite the high values of $\Rm$ in these flows, AW can undergo high dissipation in thin regions, for example in the solar corona where anomalous heating occurs \citep{davila1987_apj,singh2007_sp}. 
Understanding how AW dissipate energy and studying their nonlinear regime in controlled laboratory conditions may thus offer a convenient alternative to observations to understand these mechanisms at a fundamental level. Until now, however, only linear waves have been experimentally produced in liquid metals because of the large magnetic dissipation they undergo when $\Rm \ll 1$ and the conditions of their existence at low $\Rm$ are not understood. To address these questions, we force AW with an alternating electric current in a liquid metal in a transverse magnetic field. We provide the first mathematical derivation of a wave-bearing extension of the usual low-$\Rm$ magnetohydrodynamics (MHD) approximation to identify two linear regimes: the purely diffusive regime exists when $\NOmega$, the ratio of the oscillation period to the time scale of diffusive two-dimensionalisation by the Lorentz force, is small; the propagative regime is governed by the ratio of the forcing period to the AW propagation time scale, which we call the Jameson number $\Ja$ after \cite{jameson_1964}. 
In this regime, AW are dissipative and dispersive as they propagate more slowly where transverse velocity gradients are higher. Both regimes are recovered in the FlowCube experiment \citep{pk2014_jfm}, 
in excellent agreement with the model up to $\Ja \lesssim 0.85$ but near the $\Ja = 1$ resonance, high amplitude waves become clearly nonlinear. Hence, in electrically driving AW, we identified the purely diffusive MHD regime, the regime where linear, dispersive AW propagate, and the regime of nonlinear propagation.

	\thispagestyle{myheadings}
\end{abstract} 



\section{Introduction}
\label{sec:intro}
The main purpose of this work is to determine whether Alfv\'en waves (AW) produced in liquid metal experiments can bear relevance to those in stellar or geophysical systems. Specifically, the question is whether AW can be excited and can reach sufficiently high intensity to generate complex, possibly nonlinear, dynamics despite the high dissipation they undergo. 

In the absence of dissipation, AW are incompressible, non-dispersive waves propagating in electrically conducting fluids along a background magnetic field $\boldsymbol B_0$ at a phase velocity $V_A=B_0/\sqrt{\rho\mu}$, where $\rho$ is the fluid's density and $\mu$ the magnetic permeability of the vacuum \citep{roberts_1967,moreau1990,davidson2001,finlay2007}.  
In a medium of finite conductivity $\sigma$, they dissipate over a time scale $\tau_d=L^2/\eta$, where $\eta=(\sigma\mu)^{-1}$ is the magnetic diffusivity. Because of the very large scale $L$ of stellar and geophysical systems,  $\tau_d$ is many orders of magnitude (typically 10) greater than the propagation time scale $\tau_A=L/V_A$. 
Hence, AW propagate practically unimpeded in the very low density stellar and interstellar media, the solar wind, planetary magnetospheres but also in much denser planetary interiors, where they play an important role in energy transport and dissipation \citep{tsurutani1999_rg,nakariakov1999_sci,jault2015_tg}. The low dissipation favours large amplitudes and nonlinearities that underpin energy transfers between them. The solar wind offers a good example of Alfv\'enic turbulence, where such transfers operate across a very wide range of length scales \citep{salem2012_apj,howes2015_prsa}.
In the Sun, AW are one of the main candidates for explaining high temperatures in the solar corona \citep{grant2018_natphys,li2021_apj}. They are also commonly encountered in magnetised planetary cores, under the form of torsional AW propagating between concentric cylinders aligned with planets rotation 
\citep{bragingsky1970_ga,gillet2010_nat}, or as magneto-Coriolis waves \citep{finlay2008,gillet2022_pnas,varma2022_pepi,majumder2023_pepi}.
Unfortunately, these waves are extremely difficult to study in their natural environment. Accessibility is an obvious reason, but by no means the only one: 
observational data produced by satellites deliver limited local data. Furthermore, Alfv\'en waves compete with several other magnetomechanical oscillations, for example incompressible oscillations of solar coronal loops, or magnetoacoustic waves arising out of the medium's compressibility (see \cite{nakariakov2020_araa} for a review). 
Distinguishing AW amongst these, especially with limited observational data, poses a significant challenge. Numerical simulations are challenging too because of the extreme Reynolds and magnetic Reynolds numbers at which these systems operate.

For these reasons, producing carefully controlled Alfv\'en waves in a laboratory formed an appealing proposition to understanding their role in natural systems ever since they were first theorised by \cite{alfven_1942} in his seminal, yet remarkably simple paper. The immediate obstacle to such an endeavour arises from the $10^8$ to $10^{10}$ factor between lengthscales of experiments and natural systems, which inflicts just as drastic a reduction in the ratio $\tau_d/\tau_A$. While higher magnetic fields linearly reduce $\tau_A$, even the highest magnetic fields available to date (10 T or more) ever regain three or four orders of magnitude at best. Such is the challenge of keeping this ratio sufficiently high to observe AW, that it was named after the pioneer who made the first attempt \citep{lundquist_1949}: the Lundquist number $\Lu=V_A L/\eta$. Lundquist tried to force AW with a conducting disc oscillating across a background magnetic field in a mercury vessel and measured the intensity of these oscillations farther down the field lines. The amplitude of the oscillations was weak with a frequency dependence relatively far from the non-viscous model he tried to match it to. Lehnert's subsequent attempt \citep{lehnert_1954} was based on a similar mechanical principle. Despite improved instrumentation and control he arrived at a similar result. While Lundquist and Lehnert's experiments set milestones as the first laboratory experiments seeking to produce AW, the most convincing evidence of AW in liquid metals is due to \cite{jameson_1964}. The basis for his success was two new ideas: he showed theoretically that forcing waves electromagnetically instead of mechanically led to higher amplitudes and took advantage of the spatial inhomogeneity of the waves to place his probes at the locus where a local resonance maximised their amplitude. Jameson's AW precisely matched the linear theory but their amplitude was still too low for nonlinear effects to even be noticeable. \citet{alboussiere_2011} tried to venture in this regime by taking advantage of electromagnets delivering up to $13\,$T in a $160\,$mm diameter bore: these authors tried to produce a self-interacting wave with an electromagnetic pulse bouncing against the ends of a vessel filled with Galinstan, a eutectic alloy liquid at room temperature. Alfvén waves were produced but their decay was too fast for the reflected and incident waves to interact. Being poloidal, these waves differed from Alfv\'en's theory and previous liquid metal experiments on parallel, transversal waves. Further recent experiments focused on different aspects of linear AW:
\citet{iwai2003_mhd} extracted the signature of AW from the pressure fluctuations. Magneto-Coriolis \citep{nornberg2010_prl, schmitt2013_ejmb} and torsional AW were produced more recently in liquid metal spherical Couette experiments \citep{tigrine2019_gji}. The most extreme AW experiment to date is without doubt due to \cite{stefani2021_prl}, who produced AW within a small capsule of rubidium subjected to a $63\,$T pulsed magnetic field of $150\,$ms. In this regime, compressibility enables these authors to excite a parametric resonance between magnetoacoustic and AW, when the velocities of sound and AW coincide \citep{zaqarashvili_AA_2006}. 
This mechanism involving compressibility bears relevance to those generating heat in the solar corona and this experiment is the only liquid metal experiment to have reached a nonlinear AW regime.\\ 
These experiments clearly identified a propagative behaviour with some resemblance to AW. However, the discrepancies with theory observed by \cite{lundquist_1949} and \cite{lehnert_1954} raise the question of the nature of the waves observed and how different these may be from Alfv\'en's ideal waves. While \cite{jameson_1964} and \cite{alboussiere_2011} obtained a much better agreement with theory, they did so by better incorporating the effect of diffusion but stopped short of characterising how AW, their topology and propagation properties were affected by it. Lastly, the question of the conditions in which AW even exist in the presence of diffusion is still unexplored.

Independently, liquid metal magnetohydrodyanmics (MHD) at laboratory scale has developed since the 1960s with the common assumption that the induced magnetic field is small enough for the magnetic induction diffusion to be several orders of magnitude greater than its advection by the flow, i.e. that the magnetic Reynolds number $\Rm=UL/\eta$ is vanishingly small (except where a dynamo effect was specifically sought). 
It became common, and sometimes justified to assume that the time scale of magnetic field fluctuations is  the flow advection time scale $L/U$, i.e. 
$\partial_t \boldsymbol B\sim \mathbf u \bm\cdot\nabla \boldsymbol B$ 
(see for instance \cite{knaepen_JFM_2004}). 
This widely used assumption effectively merges the low$-\Rm$ approximation with the quasistatic MHD (QSMHD) approximation, where magnetic field fluctuations are smeared out by diffusion \citep{sarris_NumHeat_B_2006,knaepen_Ann_RFM_2008,favier_JFM_2011,sarkar_EuroJM_2019}. 
Merging these two time scales implies that Alfvén waves cannot exist at low $\Rm$, which is incorrect \citep{ennayar2021_epm} when this assumption is not justified \citep{jameson_1964,roberts_1967}.	
This context and the limited success of liquid metal experiments in producing AW 
led to the idea that low-$\Rm$ liquid metal experiments could not produce AW of relevance to their natural settings where $\Lu\gg1$ and $\Rm\gg1$. Plasmas, by contrast, soon appeared as an alternative to liquid metals due to their naturally high Lundquist numbers, especially when plasma technology for nuclear fusion emerged in the 1950s. The first  indisputable evidence of AW was indeed obtained from a measure of wave velocity in a plasma \citep{allen1959_prl}, followed by further experiments \citep{wilcox1961_pf,jephcott1962_jfm,woods1962_jfm}, and the generation of linear interferences between AW \citep{gekelman1997_jgr}. The first AW nonlinearities in plasmas were, however, obtained much more recently: the observation of  parametric instabilities and nonlinear transfers between AW \citep{carter2006_prl,dorfman2016_prl} achieved an important step towards Alfvénic turbulence in the laboratory \citep{howes2012_prl, howes2013_pp}. While the compressibility of plasmas makes it more difficult to isolate AW from other waves, for example magnetoacoustic waves \citep{dorfman2013_prl}, it also makes them relevant to the solar corona and the solar wind. Plasmas also pose serious challenges in terms of metrology and require much heavier technological environments than liquid metals. Liquid metals, by contrast are incompressible, very dense and bear close similarities with planetary cores, but current experiments with them are scarce and mostly target the linear regime. 

As a result, the current state of understanding of AW is relatively limited, considering they were discovered over 80 years ago. To this date there have been no experiments reproducing Alfvénic turbulence or any which are able to reproduce the complex nonlinear effects taking place in solar or geophysical systems. Even deriving reliable laws for their reflections against walls poses considerable challenges \citep{schaeffer2012_jgi}. 
Yet, evidence of their role in stellar and planetary interiors accumulates in simulations, and observations \citep{tomczyk2007_sci,gillet2010_nat,grant2018_natphys,nakariakov2020_araa,gillet2022_pnas,li2021_apj,schwaiger2024_pepi}. A new form of helioseismology and planetary seismology using AW is even emerging as powerful means of probing the interior of the Sun and Jupiter \citep{hanasoge2012_prl,hori2023_natastro}. It is also becoming increasingly clear that wherever AW play a role, they do so through dissipation and nonlinearity, rather than in a regime of ideal linear AW. In the solar corona, thin dissipation layers may be the missing heat source that could explain high temperatures there \citep{grant2018_natphys}, and high dissipation occurs at small scales energised by nonlinear transfer from larger scales \citep{davila1987_apj}. In spherical shells representing planetary interiors, dissipation layers are required to obtain a correct solution describing the propagation of quasi-two-dimensional torsional AW \citep{luo2022_prsa}.
These examples illustrate that the regions where AW play the most crucial role are much smaller than the planetary or solar scales, so at these scales, the local Lundquist number may fall in a range accessible to liquid metal experiments $1\lesssim\Lu\lesssim 10^2$ \citep{cattell1996_jgr,singh2007_sp}. 
The dissipative behaviour of liquid metal experiments may therefore not be as irrelevant to astrophysical and geophysical problems as the staggering values of $\Lu$ in these problems suggest. At the same time, liquid metal technology has made strides since Jameson's success: experiments \citep{klein_potherat_2010, baker_inverse_2018} conducted in very high magnetic fields (up to 10 T and rising) now offer extensive flow mapping based on ultrasound velocimetry \citep{brito2001_ef, franke2010_fmi} and low-noise high-precision electric potential velocimetry (EPV) \citep{kljukin_direct_1998,frank2001_pf,baker_controlling_2017}.\\ 
The role of dissipation and nonlinearities, the mounting need to understand complex AW and the availability of these technologies prompts us to seek the conditions in which AW can be obtained at low$-\Rm$ and whether these can be obtained sufficiently far from the QSMHD regime for AW to incur nonlinear effects. To do this, we take advantage of electric forcing and transverse inhomogeneity as \citet{jameson_1964} did, and implement these ideas in high magnetic fields by adapting the FlowCube device \citep{klein_potherat_2010,pk2014_jfm,baker_controlling_2017,baker_inverse_2018,pk2017_prf}, which we previously developed to study MHD turbulence. We seek to answer the following key questions.
\begin{enumerate}
\item In which conditions do AW propagate at low $\Rm$, despite diffusion?
\item How do dissipative AW differ from the ideal non-dissipative, non-dispersive AW, especially where they are not homogeneous in planes normal to the field?
\item Can a nonlinear regime of AW be reached in liquid metals at low $\Rm$?
\end{enumerate}
We first revisit the low$-Rm$ approximation, to specifically allow wave propagation, i.e. outside the QSMHD regime ($\S$ \ref{sec:theory}).
Based on this approximation, a semi-analytical model for the propagation of linear AW in a plane channel normal to a background magnetic field is derived. Using this model, we analyse the flow in a channel forced by an alternating current (AC) injected at a localised electrode embedded in one of the walls, to identify the diffusive and the propagative regimes ($\S$ \ref{sec:ac_waves}). 
Then, a similar flow is experimentally generated with FlowCube whose principle and electric potential measurement system are summarised in $\S$ \ref{sec:exp_approach}. By comparing theory and experiments, we finally identify the diffusive and propagative regimes in the experiment, and seek nonlinearities where discrepancies between model and experiment arise ($\S$ \ref{sec:exp_results}).

\section{Linear theory of confined Alfv\'en waves}\label{sec:theory}
\subsection{General configuration and governing equations}
\begin{figure}[h!]
	\centering
	\includegraphics[width=0.5\linewidth]{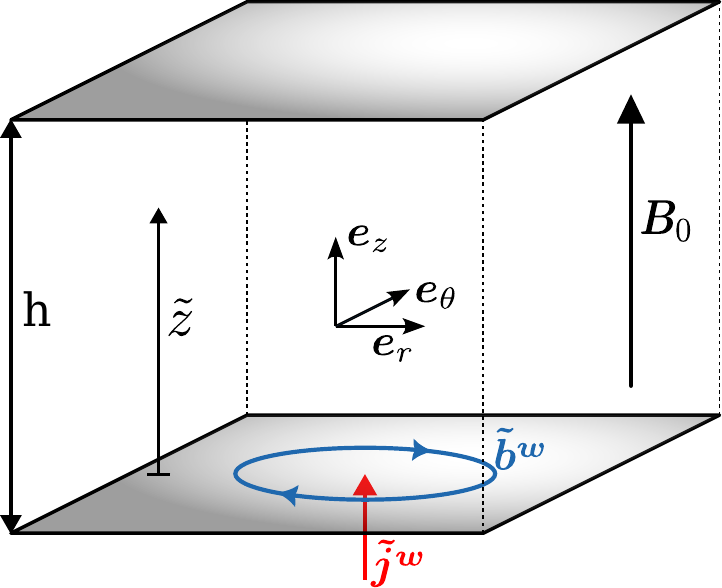}
	\caption{{Sketch of the general configuration, where Alfvén waves confined between two horizontal walls spaced $h$ apart evolve in the electrically conducting incompressible fluid subjected to a homogeneous, static and axial magnetic field $\bm B_0= B_0 \unitz$. An axial AC current density $\bm{\tilde j}^w$ is injected at the bottom wall, which can be expressed in term of the magnetic disturbance $\bm{\tilde{b}}^w$ by means of Ampère's law. The top wall is electrically insulated. Both walls are solid, no-slip and impermeable. 
}}
	\label{fig:sketch_general_geometry}
\end{figure}

We consider a channel of height $h$ (figure \ref{fig:sketch_general_geometry}) filled with an electrically conducting incompressible fluid of density $\rho$, kinematic viscosity $\nu$ and electric conductivity $\sigma$ and subjected to an axial, homogeneous and static magnetic field $\boldsymbol{B_0} = B_0\,\boldsymbol{e_z}$. The domain is bounded by two horizontal solid, impermeable and electrically insulating walls {at} $\tilde{z} = 0$ and $\tilde{z}= h$, thereafter referred to as Hartmann walls as they are normal to the magnetic field. 
The flow is forced at the bottom wall by injecting an axial AC current density $\tilde{\bm j}_z\left(\tilde{r},{\theta},\tilde{z}=0,\tilde{t}\right)= \tilde{j}^w\left(\tilde{r},{\theta}\right) \,\cos\left(\omega\,\tilde{t}\right)\unitz$ of angular frequency $\omega$ and 
magnitude $\tilde{j}^{w}$. The amplitude of the total current injected at the bottom wall is $I_0$.
Normalising distances by $h$, time by $2\pi/\omega$, velocity by $u_0$, magnetic fields by $B_0$, current density by $j_0= I_0/h^2$ {and pressure by $\rho u_0^2$}, the governing equations for the velocity $\boldsymbol u$, pressure $p$ and magnetic disturbance to the external field $\boldsymbol b= \boldsymbol B - B_0\boldsymbol e_z$ are the Navier-Stokes equations, the induction equation, as well as the conservation of mass and magnetic flux,
\begin{align}
\Rv \, \partial_t \boldsymbol u + \Rey\lbrace \boldsymbol u \bm\cdot \bm\nabla \boldsymbol u + {\bm\nabla p} \rbrace &= \bm\Delta \boldsymbol u + \Ha^2\,\Rm^{-1}\lbrace\partial_z \boldsymbol{b} + \boldsymbol b\bm\cdot\bm\nabla\boldsymbol b\rbrace, \label{eq:ns_adm}\\
\Rn \partial_t \boldsymbol b  &=  \bm\Delta\boldsymbol b - \Rm\lbrace \boldsymbol u\bm\cdot \bm\nabla \boldsymbol b - \boldsymbol b\bm\cdot \bm\nabla \boldsymbol u - \partial_z \boldsymbol u \rbrace, \label{eq:induction_adm}\\
\bm\nabla\bm\cdot \boldsymbol{u} &= 0,\label{eq:mass}\\
\bm\nabla\bm\cdot \boldsymbol{b} &= 0,\label{eq:gauss}
\end{align}
{where $\bm\Delta$ is the vectorial Laplacian operator}.
The five non-dimensional numbers governing this system are, respectively, the Reynolds number, the magnetic Reynolds number, the Hartmann number and the screen parameters for viscous and magnetic diffusion,
\begin{equation}
\Rey = \frac{u_0 h}{\nu}, \ \ \Rm=\frac{u_0h}{\eta},  \ \ \Ha= B_0\,h\sqrt{\frac{\sigma}{\rho\nu}}, \ \  \Rv = \frac{\omega h^2}{2\pi\nu} \ \ \Rn = \frac{\omega h^2}{2\pi\eta} , 
\end{equation}
where $\eta = \left(\sigma\mu_0\right)^{-1}$ is the magnetic diffusivity.

Here, $\Rm$ measures the ratio of magnetic field advection by the flow to magnetic field diffusion {while $\Rey$ measures the ratio between inertia and viscous forces.} The square of the Hartmann number $\Ha$ measures the ratio of Lorentz to viscous forces. Finally, $\Rv$ (respectively, $\Rn$) measures the square ratio of the viscous diffusive (respectively, resistive) penetration depth of boundary oscillations into the domain to the domain's size (see for example \cite{batchelor_1967}). 

{So far, the characteristic velocity $u_0$ is not defined. 
However, driving flows with a current $I_0$ injected at a single electrode induce a circulation $\Gamma = I_0/[2\pi\left(\rho\sigma\nu\right)^{1/2}]$, which together with lengthscale $h$ provides the usual velocity scale for electrically driven flows and the corresponding Reynolds number \citep{sommeria1988_jfm},
\begin{equation}
u_0=\Gamma/h, \qquad \Rey_0 = I_0/[2\pi\nu\left(\sigma\rho\nu\right)^{1/2}].
\label{eq:u0}
\end{equation}
The quantity $\Rey_0$ thus provides a non-dimensional measure of the forcing intensity \citep{klein_potherat_2010}. 
Additionally, the magnetic Reynolds number based on $u_0$, $\Rm_0$, is also determined by this choice, 
since the ratio $\Rm_0/\Rey_0 = \Rn/\Rv = \Pm$ is the magnetic Prandtl number and is fixed for a given choice of fluid.  
It should be noted that since the expression of $\Gamma$ accounts only for the Lorentz force due to the injected current and dissipation in the Hartmann layers, the scale $u_0$ may significantly overestimate the actual velocities in the fluid. Hence, it should be better understood as a measure of the forcing. Reynolds and magnetic Reynolds numbers based on actual velocities (defined in $\S$ \ref{sb_sec:measured_Re_Rm}) are discussed in $\S$ \ref{sec:exp_results}. 
}

The kinematic boundary conditions at the top and bottom Hartmann walls 
are no-slip impermeable,
\begin{equation}
    {\boldsymbol u(r,\theta,z=0,t)=\boldsymbol u(r,\theta, z=1,t)=0.}
\label{eq:bc_u}
\end{equation}
At the bottom wall ($z=0$) the normal electric current is imposed and the top wall ($z=1$) is electrically insulating. These conditions are expressed in terms of the magnetic disturbance $\boldsymbol b$ by means of Ampere's law, \\
\begin{eqnarray}
\boldsymbol{\bm\nabla} \times {\boldsymbol{b}}({r},{\theta},z=0,{t}) \bm\cdot \boldsymbol{e}_z &=&j^w{\left(r,\theta\right)\,\cos\left(t\right)},\label{eq:bc_bbot}\\
\boldsymbol{\bm\nabla} \times {\boldsymbol{b}}({r},{\theta},z=1,{t}) \bm\cdot \boldsymbol{e}_z &=&0.
\label{eq:bc_btop}
\end{eqnarray}
Integration in the plane of each Hartmann wall, respectively, leads to inhomogeneous and homogeneous conditions for $\boldsymbol b$ at the bottom and top walls, 
\begin{equation}
    \boldsymbol b_\perp({r},{\theta},z=0,{t})=\boldsymbol b^w(r,{\theta})\,{\cos\left(t\right)}, \qquad \boldsymbol b({r},{\theta},z=1,{t})=0,
	\label{eq:bc_b}
\end{equation}
where the subscript $_\perp$ stands for projection in the horizontal plane {$(r,\theta)$} and $\boldsymbol b^w$ is uniquely defined by the choice of $j^w$.

\subsection{A propagative low-$\Rm$ approximation
\label{sec:lowrm}}
We start by simplifying the governing equations in the limit $\Rm\rightarrow0$ 
in way suitable to describe the propagation of waves in liquid metals.
{In the low-$\Rm$ approximation, the physical quantities $\bm b$ and $\bm u$ are expanded in powers of $\Rm$.
Thus, the induction equation at the leading order, $ O (\Rm^0)$, readily implies}
\begin{equation}
\Rn \partial_t \boldsymbol b=\bm\Delta \boldsymbol b.
\end{equation}
Indeed, for the fluid motion to actually induce a magnetic field, the transport term and the magnetic diffusion must balance which implies  $\bm\Delta\boldsymbol b=O(\Rm)$ and therefore $\boldsymbol b=O(\Rm)$ \citep{roberts_1967}. 
{Since $\boldsymbol u=O(\Rm^0)$, this implies that $b/u=O(\Rm)$ so that forcing AW electromagnetically requires forcing amplitudes $\Rm$ times smaller than forcing them mechanically. Hence, as noted by \citet{jameson_1964}, AW are much more efficiently forced electromagnetically than using the sort of mechanical forcing of the early experiments of \citet{lundquist_1949} and \citet{lehnert_1954}. Denoting $\widetilde{\boldsymbol{b}}=\Rm^{-1}\boldsymbol{b}$ and} {keeping $O(\Rm^0)$ terms in (\ref{eq:ns_adm}) and all highest remaining terms, i.e. of order $O(\Rm)$, in the induction equation (\Ref{eq:induction_adm}) yields
}
\begin{align}
\left(\Rv \, \partial_t \boldsymbol - \bm\Delta \right) \boldsymbol u&=  \Ha^2\,\partial_z  \rmb, \label{eq:linear_ns_adm}\\
\left(\Rn \, \partial_t \boldsymbol - \bm\Delta \right) \rmb&= \partial_z \boldsymbol u. \label{eq:linear_induction_adm}
\end{align}

Importantly, {since these equations result from an asymptotic expansion in $\Rm$}, $\Rm$ disappears in the linearised low-$\Rm$ approximation but three governing non-dimensional numbers are left: the Hartmann number $\Ha$ and the two screen parameters $\Rn$ and $\Rv$. 
Eliminating either $\boldsymbol b$ or $\boldsymbol u$ reveals that both variables obey the same formal equation 
\begin{equation}
\left(\left(\Rv\,\partial_t \boldsymbol - \bm\Delta \right)\left(\Rn \, \partial_t \boldsymbol - \bm\Delta \right)- \Ha^2 \partial^2_{zz} \right) \lbrace \boldsymbol{u}, \rmb \rbrace = 0.\label{eq:governing_adm}
\end{equation} 
{{Equations (\ref{eq:linear_ns_adm})-(\ref{eq:linear_induction_adm}) or equation (\ref{eq:governing_adm}) are the mathematical expression of the propagative low-$\Rm$ approximation in the limit $Re\rightarrow0$. They are different from the QSMHD approximation discussed in \S \ref{sec:qsmhd}, which requires the additional assumption that $\Rn\rightarrow0$.} The equations for the low-$Rm$ approximation for arbitrary $Re$ are obtained simply by retaining the nonlinear and pressure terms 
$\Rey\lbrace \boldsymbol u\bm\cdot \bm\nabla \boldsymbol u + {\bm\nabla p} \rbrace$
in the left-hand side of (\ref{eq:linear_ns_adm}). 
Several authors used a MHD equation resembling these. The equations of Lehnert
(\citeyear{lehnert1955_qam}) and Moffatt (\citeyear{moffatt1967_jfm}) include the $\partial_t\boldsymbol{b}$ term but are missing the $\boldsymbol{u}\bm\cdot\nabla\boldsymbol{b}$ term in the induction equation, and the nonlinear term in the momentum equation.
While these are derived dimensionally, they are not obtained directly as an asymptotic form of the full MHD equations in the limit $\Rm\rightarrow0$ and only apply to infinitesimal magnetic field and velocity perturbations of a background state with a constant magnetic fields and no flow.  
\citet{knaepen_JFM_2004} started from the dimensional full MHD equations and dropped the $\boldsymbol{u}\bm\cdot\nabla\boldsymbol{b}$ term in the induction equation but kept the full nonlinear terms in the Navier-Stokes equations. These equations, termed quasilinear are effectively the dimensional form of the propagative low-$Rm$ approximation.
They were justified empirically by comparing numerical simulations of the full MHD equations and the QSMHD equations but not analytically derived from the full MHD equations. 
Here, introducing a separate time scale for the terms with time derivative enabled us to derive the propagative low-$\Rm$ as an asymptotic limit of the full MHD equations for the first time. This provides a rigorous justification to the equations used by \citet{knaepen_JFM_2004}'s. Indeed the numerical simulations conducted by these authors confirm the effectiveness of the propagative low-$\Rm$ approximation for $R\!m\lesssim1$.}

Up to this point, boundary conditions have only been specified at the top and bottom walls, but the conditions at the lateral boundaries have remained unspecified. Let us now assume that these allow for a solution of Eq (\ref{eq:governing_adm}) to be found by separation of variables under the form
{
\begin{equation}
\{\boldsymbol{u}, \rmb\}\left(r,\theta,z,t\right)=\{\boldsymbol{U}^{\perp} \left(r,\theta,z\right) \boldsymbol{U}^{z} \left(r,\theta,z\right), \boldsymbol B^{\perp}\left(r,\theta,z\right)\boldsymbol B^{z}\left(r,\theta,z\right)\}\,\exp\left({j\epsilon t}\right) + \mathrm{c.c.},\label{eq:solution_base_eigenfunction}
\end{equation}
}
where $\epsilon = \pm 1$ {and $j$ is the imaginary unit}. Here, $\left\{\boldsymbol{U}^{\perp},\,\boldsymbol{B}^{\perp}\right\}$ and $\left\{\boldsymbol{U}^{z},\,\boldsymbol{B}^{z}\right\}$   
are the eigenfunctions of the Sturm\textendash Liouville problems, respectively, associated with the directions perpendicular and parallel with the magnetic field directions, with respective eigenvalues $(-\lambda_\perp,-\lambda_z)\in\mathbb C^2$,
\begin{eqnarray}
\left(\Delta_{\perp} + \lambda_{\perp}\right) \{{\boldsymbol{U}^{\perp}},{\boldsymbol{B}^{\perp}}\}&=&0, \label{eq:Helmholtz_t_laplacian}\\
\left(\partial_{zz}^2 + \lambda_{z}\right) \{{\boldsymbol{U}^{z}},{\boldsymbol{B}^{z}}\}&=&0 \label{eq:Helmholtz_t_laplacian_z},
\end{eqnarray} 
with $\Delta_\perp=\bm\Delta-\partial^2_{zz}$.
This decomposition leads to the dispersion relation,
\begin{align}\label{eq:disp_relation}
\lambda_z^2 + \left( \Ha^2 + 2\lambda_{\perp} +\epsilon j \left(\Rv+\Rn\right)\right) \lambda_z 
- \Rv \Rn + \lambda_{\perp}^2 + \epsilon j \left(\Rv+\Rn\right) \lambda_{\perp} =0.
\end{align}
The boundary conditions at $z=0$ and $z=1$ impose that $\boldsymbol{U}^{z}$ and $\boldsymbol{B}^{z}$ {be} of the form {$C\exp\{j(\kappa+js) z\}$}, where $\lambda_z=(\kappa+js)^2$ incorporates real wavenumber $\kappa$ and spatial 
attenuation $s$. The eigenvalues $\lambda_\perp$ are determined by the geometry and boundary conditions in the horizontal plane. These are left unspecified for now, but we shall simply assume that $\lambda_\perp=\kappa_\perp^2>0$ to cover the most common cases such as rectangular, periodic or axisymmetric domains. This choice enables us to introduce a real transverse wavenumber $\kappa_\perp\in\mathbb R$.
Under these assumptions, the dispersion relation admits four solutions $\lambda_{z,m} = \pm\left( \kappa_{m} + j s_m\right)^2$, where $ m \in\left\lbrace 1, 2 \right\rbrace$, is expressed as
\begin{align}
\kappa_{m}&= \pm \frac{1}{2}\bigg[ -\Ha^2 - 2\,\kappa_{\perp}^2 + \epsilon_m\,a + \left\lbrace \left(-\Ha^2 - 2\,\kappa_{\perp}^2 + \epsilon_m\,a\right)^2 + \Big(-\Rv - \Rn + \epsilon_m\,b\Big)^2\right\rbrace^{1/2} \bigg]^{1/2}, \label{eq:kz_i}\\
s_m&= \pm \epsilon\frac{1}{2}\bigg[ \Ha^2 + 2\,\kappa_{\perp}^2 - \epsilon_m\,a + \left\lbrace \left(-\Ha^2 - 2\,\kappa_{\perp}^2 + \epsilon_m\,a\right)^2 + \Big(-\Rv - \Rn + \epsilon_m\,b\Big)^2\right\rbrace^{1/2} \bigg]^{1/2}, \label{eq:sz_i}
\end{align}
with
\begin{align}
a =& \frac{1}{\sqrt{2}}\bigg[\Ha^4 + 4\kappa_\perp^2 \Ha^2 - \left(\Rv - \Rn\right)^2  + \notag \\
+ & \left\lbrace \left[ \Ha^4 + 4\kappa_\perp^2 \Ha^2 - \left(\Rv - \Rn\right)^2\right]^2 + 4\Ha^4 \left( \Rv + \Rn\right)^2 \right\rbrace^{1/2}\bigg]^{1/2},\label{eq:a}\\
\widetilde{b} =& \frac{1}{\sqrt{2}}\bigg[-\Ha^4 - 4\kappa_\perp^2 \Ha^2 + \left(\Rv - \Rn\right)^2  + \notag \\
+ & \left\lbrace \left[ \Ha^4 + 4\kappa_\perp^2 \Ha^2 - \left(\Rv - \Rn\right)^2\right]^2 + 4\Ha^4 \left( \Rv + \Rn\right)^2 \right\rbrace^{1/2}\bigg]^{1/2},\label{eq:b}
\end{align}
and $\epsilon_m = \left(-1\right)^m$. The two families of solutions $m=1$ and $m=2$ have very different damping and propagation properties. Solutions from the first family (subscript 1) decay very fast in the $z-$direction, as the spatial decay rate $s_1$ is always greater than the wavenumber $\kappa_{1}$, and so the corresponding profile mostly follows an exponential decay away from the top and bottom boundaries, without achieving a complete oscillation. Physically, this mode represents the Hartmann boundary layers that develop along the bottom and top walls and we shall refer to it as Hartmann mode for this reason. As such, $s_1\sim \Ha$ in the limit $\Ha\rightarrow\infty$, keeping $\kappa_\perp\ll \Ha$. 

In contrast, solutions from the second family (subscript 2) are much less attenuated in the $z-$direction. Depending on the value of parameters $\Ha$ and $\kappa_{\perp}$, a range of values of $\Rn$ may exist such that $s_2 < \kappa_{2}$. In other words, an oscillatory solution develops into the fluid layer, which can describe the propagation of a wave. Hence, we shall refer to this mode as the {Alfv\'en mode}. 
The properties of the Alfvén mode are best illustrated in the diffusionless limit, where the acceleration term (with the time derivative) and the Lorentz force (last term of Eq. (\ref{eq:governing_adm})) balance each other, as the diffusive terms become negligible compared with them. This regime is achieved in the limit where $\Rv\rightarrow\infty$ and $\Rn\rightarrow\infty$ keeping $\Ha^2/(\Rv \Rn)$ finite for the Lorentz force to remain finite.  This number 
characterises waves {driven at a} specific frequency $\omega$ such as in {Lundquist}, Lehnert and Jameson's experiments. 
It can be expressed as 
the inverse of a Lundquist number based on the time scale of the oscillation instead of the magnetic diffusion time scale. 
However, since \cite{jameson_1964} was the first to have successfully produced strong AW resonances, precisely by adjusting the forcing frequency, we propose to name this number after him and define the Jameson number as $\Ja = \left(\Rn \Rv\right)^{1/2}/\Ha = \Rn/(V_Ah/\eta)=\Rn/\Lu$, where $V_A=B_0/(\rho\mu_0)^{1/2}$ is the Alfv\'en velocity. {When $\Ha^2 \rightarrow \infty$ but keeping $\Ha^2/\left(\Rn \Rv\right)$  constant and finite,} Eq. (\ref{eq:governing_adm}) then reduces to the purely hyperbolic equation for the propagation of these waves in an ideal medium,
\begin{align}
	\left(\partial^2_{tt}- \Ja^{-2}\partial_{zz}^2\right) \lbrace \boldsymbol{u}, \rmb \rbrace = 0.\label{eq:governing_adm_alfven}
\end{align}
In the absence of viscous diffusion, the governing equations drop to second order and the no-slip conditions at the top and bottom boundaries need not be satisfied. For an AC injected current $j^w$ with sinusoidal waveform, a sinusoidal wave propagates through the layer, with wavenumber found as the asymptotic value of $\kappa_2$ in this limit,
\begin{align}
	\kappa_{2}\sim\pm\Ja=\pm\frac{\left(\Rn\Rv\right)}{\Ha}^{1/2}.
\end{align}
Since the influence of the horizontal geometry (through $\kappa_\perp$) disappears in the diffusionless limit, diffusionless AW are non-dispersive. Hence, for a flow forced by imposing a boundary condition at one of the Hartmann walls, this boundary condition simply propagates uniformly and without dispersion along $z$ at speed $V_A$. This also illustrates that the dispersive nature of the waves propagating outside the non-dissipative regime stems from the magnetic and viscous dissipation. 
\subsection{The QSMHD limit
\label{sec:qsmhd}}
While the set of governing equations (\ref{eq:linear_ns_adm}-\ref{eq:linear_induction_adm}) potentially supports waves, it only does so when the magnetic field oscillations are not fully damped by magnetic diffusion. By contrast, the waveless regime where {waves} are overdamped takes place in the QSMHD limit, where magnetic diffusion acts much faster than the time scale of
{ the induced magnetic field fluctuations, i.e. when $\Rn\rightarrow0$.} 
In the Navier-Stokes equation, on the other hand, since no assumption is made on $\Ha$ or $\Rv$, $\Rv\,\Rm=O(\Rm)$, $\Ha^2 \partial_zb\sim \Ha^2\,\Rm=O(\Rm)$ so all terms in Eq. (\ref{eq:linear_ns_adm}) are $O(\Rm)$ and must be retained. 

In other words, while the resistive screen parameter $\Rn$ must be $0$ in the QSMHD limit, the viscous screen parameter $\Rv$ and the Hartmann number $\Ha$ may retain finite values. Since, $\Rv$ remains finite whilst $\Rn$ vanishes, this approximation requires that $P\!m=\Rn/\Rv\rightarrow0$, and so applies to MHD flows of liquid metals or conducting electrolytes \citep{andreev2013_jfm,apbds2016_rsi,mph2020_ef} but not necessarily to plasmas. In the end, the linearised low-$\Rm$ QSMHD equations take the form 
\begin{eqnarray}
	\left(\Rv \, \partial_t \boldsymbol - \bm\Delta \right)\boldsymbol u&=&  \Ha^2\,\partial_z \rmb, \label{eq:linear_ns_adm_qs}\\
	\bm\Delta \widetilde{\boldsymbol b} &=&  -\partial_z \boldsymbol u, \label{eq:linear_induction_adm_qs}
\end{eqnarray}
and the governing equation for  $\boldsymbol u$ and $\boldsymbol b$ simplifies to
\begin{equation}
\left(\left(\Rv \, \partial_t \boldsymbol - \bm\Delta \right)\bm\Delta+ \Ha^2 \partial_{zz}^2 \right) \lbrace \boldsymbol{u},\, \rmb \rbrace = 0,\label{eq:governing_adm_qs}
\end{equation} 
with the following dispersion relation
\begin{align}\label{eq:disp_relation_QS}
\lambda_z^2 + \left( \Ha^2 +2 \,\lambda_{\perp} +\epsilon j \Rv\right) \lambda_z
	+\epsilon j \Rv\lambda_\perp+\lambda_{\perp}^2=0.
\end{align}
Formally, Eq. (\ref{eq:governing_adm_qs}) expresses the eigenvalue problem for the dissipation operator $\bm\Delta-\Ha^2\bm\Delta^{-1}\partial_{zz}^2$, with $\epsilon j\Rv$ as the eigenvalue \citep{pa2003_pf,pa2006_pf}. The corresponding eigenfunctions offer a minimal basis for the representation of MHD flows in the QSMHD approximation \citep{dyp2009_tcfd,pd2010_jfm,kop2015_jcp,pko2015_jfm}. In the QSMHD approximation, the Lorentz force acts to diffuse momentum of transverse lengthscale $l_\perp$ along the magnetic field over distance $l_z$  in time scale $\tau_{2D}=(\rho/(\sigma B^2))(l_z/l_\perp)^2$ \citep{sommeria_why_1982}. Structures of sufficiently large scale for this process to overcome viscous dissipation and inertia become quasi-two-dimensional \citep{klein_potherat_2010,pk2014_jfm,baker_inverse_2018}.
{Since} waves do not propagate in the quasi-static limit, comparing solutions from the QSMHD equations with {those of the} propagative low$-\Rm$ equations derived in $\S$ \ref{sec:lowrm} provides us with an effective way to detect wave propagation.
\subsection{Choice of the induction time scale}
{The time scale of the induction term $\partial_t \boldsymbol b$ in the induction equation (\ref{eq:induction_adm}) is crucial and deserves a more detailed discussion. Here, this induction time scale $\tau_b$ is set by the frequency of the electric forcing such that $\tau_b=2\pi/\omega$. Formally, this introduces a time scale that is independent from other flow time scales and yields two non-dimensional parameters $\Rv$ and $\Rn$ built on its ratio to the viscous and ohmic dissipation time scales. 
The advantage of this approach is that the equations do not suffer from any \emph{a priori} assumption on that time scale, so in practice, $\tau_b$ can be varied arbitrary and represents any process controlling the induction.  In this sense this is the most general non-dimensional form of the induction equation. This also means that varying the frequency of the electric forcing offers a practical means of controlling the induction independently of other processes. The laws obtained with this approach can then be applied to particular cases where the induction time scale is controlled by a specific process, be it advection, convection \citep{roberts2000_jfm,deguchi2020_jfm} or anything else, simply by replacing $2\pi/\omega$ by the relevant time scale.\\
Flows that are either not externally forced, or forced sufficiently slowly for the forcing time scale $2\pi/\omega$ to be much greater than other flow time scales  are a very common and important example. Alfv\'en waves may then be produced by fluid motion itself (i.e. by advection), or occur as a result of a natural resonance. 
This leaves two possible time scales for the induction: the advective time scale $\tau_u=h/U$ or the natural time scale of AW $\tau_a=h/V_a$. 
This choice is decided by the Alfv\'en number $\Al =\tau_a/\tau_u$. If $\Al\gg1$, the fastest time scale is set by advection, so $\tau_b=\tau_u$ and $R_\eta=\Rm$, so that for  $\Rm\ll1$, the induction term becomes $O(\Rm^2)$ and the low-$\Rm$ approximation reverts to the quasistatic (QS) approximation for which no waves exist. 
If on the other hand $\Al \ll 1$, 
then $\tau_b=\tau_a$, $\Rn=\Lu$ and waves can exist only if $\Lu\gg1$ (typically $\Lu\gtrsim10$), i.e. if they overcome ohmic dissipation. In this case, the low-$\Rm$ approximation does not revert to the QS approximation and may support waves.\\
Yet, by far the most common choice for $\tau_b$ at $\Rm\ll1$ is $\tau_u = h/u_0$; see for example \cite{knaepen_JFM_2004,sarris_NumHeat_B_2006,knaepen_Ann_RFM_2008,sarkar_EuroJM_2019}. From the discussion above, this choice is only justified when $\Al\gg1$. When $\Al\lesssim1$, setting $\tau_b=\tau_u$ artificially eliminates waves from problems where they may exist. It amounts to forcibly replacing the low-$\Rm$ approximation, which may support waves, by the QS approximation, which cannot.\\
Finally, it is noteworthy that the condition on the Alfv\'en number expresses that waves have to overcome advection to propagate too, and not just dissipation. Such a transition between advective and propagative regimes has been extensively studied for inertial waves in flows with background rotation. In particular, the transition exhibits a wavelength dependence that may result in waves being confined to specific regions of the energy spectrum in turbulent flows \citep{dickinson1978_pf,dickinson1983_jfm,yarom2014_natphys,brons2020_jfm1,brons2020_jfm2}}
\section{Electrically driven waves \label{sec:ac_waves}}
\subsection{Wave driven by injecting current with a single electrode}\label{sec:theory_1elec}
\begin{figure}[h]
	\centering
	\includegraphics[width=0.6\linewidth]{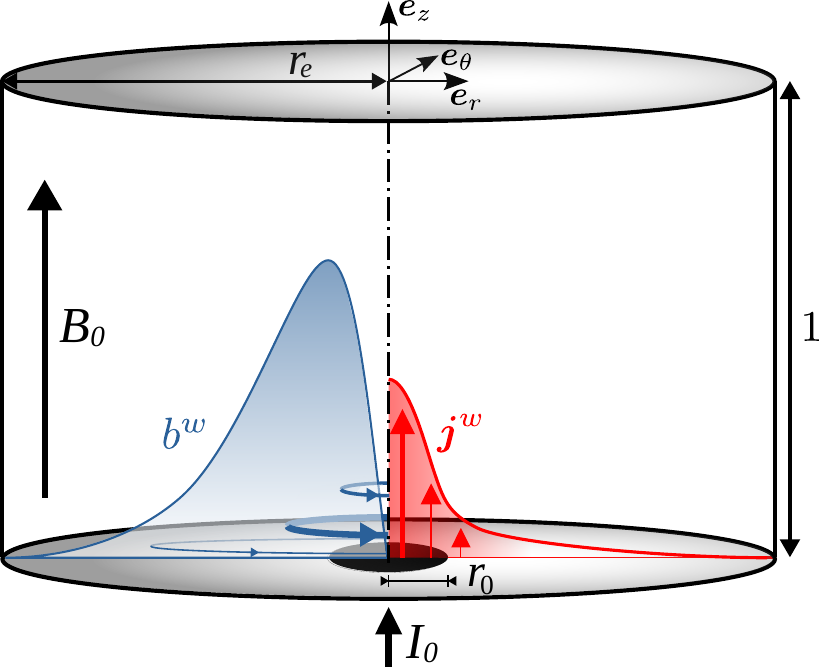}
	\caption{{Sketch of the axisymmetric geometry considered for a wave driven by injecting current with a single electrode. A cylindrical channel of radius $r_e$ closed by two horizontal, solid, impermeable and electrically insulating walls is filled with an electrically conducting incompressible fluid and subjected to a homogeneous, static and axial magnetic field $\bm B_0= B_0 \unitz$. All distances are normalised by the height of the channel. An electrode of radius $r_0$ injecting a current $I_0$ is placed flush with the bottom wall. The bottom boundary conditions on the current density and the magnetic disturbances are both represented. The right-hand side shows the radial distribution of the axial current injected by the electrode and the left-hand side shows the distribution of the azimuthal magnetic perturbation induced by the axial current. The current injected by the electrode escapes radially at infinity through the side wall.
}
}
	\label{fig:sketch_model_1elec}
\end{figure}
We now turn to the more specific case where the flow is forced by injecting an electric current at one 
{electrode embedded in the bottom wall. This is a simplified representation of the experimental device presented in $\S$ \ref{sec:flowcube}, in which an array of four electrodes is used instead. To find the conditions for Alfvén waves to emerge and identify propagative and diffusive processes, the solution is sought both in the QSMHD and in the propagative low-$\Rm$ approximations.
}

We first consider a single electrode, located at the centre of a cylindrical container of non-dimensional radius $r_e$. In the actual experiment, the current is fully localised within the radius of the electrode (typically 0.5 mm) and drops abruptly outside it. Mathematically, this would impose a discontinuity in $j^w$. To circumvent the numerical issues 
that would ensue, we therefore model the current injected at a single electrode located at $r=0$ by a sharp enough 
Gaussian distribution, on the basis that the impact of this change on the flow is limited, at least in the QSMHD limit \citep{baker_dimensionality_2015},
\begin{equation}
    \boldsymbol j^w(r)=\frac{1}{\pi\,r_0^2}\,\mathrm{exp}(-(r/r_0)^2)\,\boldsymbol e_z,
	\label{eq:jw_gauss}
\end{equation}
{where $r_0 = \tilde{r}_0/h$ is the dimensionless radius of the electrode}. The polar frame of reference ($\boldsymbol e_r$,$\boldsymbol e_\theta$) is centred on the electrode. Assuming that the magnetic perturbation is axisymmetric ($\partial_\theta = 0$) and the azimuthal magnetic perturbation vanishes at $r=0$, the Biot\textendash Savart law yields the inhomogeneous Dirichlet boundary condition for the magnetic perturbation (figure \ref{fig:sketch_model_1elec}), 
\begin{equation}\label{eq:modelled_magnetic_forcing}
\rmb_\perp \left(r,z=0,t\right)=b_{\theta}^w\left(r\right)\cos(t)\,\boldsymbol e_\theta = \left(2\pi\right)^{-1}\,\frac{1-\exp\left(-r^2/{r_0}^2\right)}{r}\cos \left(t\right) \boldsymbol e_\theta.
\end{equation}
In the general case, a similar condition $\boldsymbol u_\perp(r,z=0,t)=u^w_{\theta}(r,z=0,t)\boldsymbol e_\theta$ can be imposed on the velocity at the bottom wall to force the flow with a moving wall, as \cite{lundquist_1949} did experimentally. Here, for the purpose of modelling the experimental set-up (see $\S$ \ref{sec:exp_approach}), both Hartmann walls are considered as fixed, impermeable, with no slip solid walls and the top wall is assumed electrically insulated, 
\begin{align}\label{eq:bc_u_b_top}
\boldsymbol u_\perp(r,z=1,t)=\boldsymbol u_\perp(r,z=1,t)=0,  \qquad \rmb_\perp(r,z=1,t)=0.
\end{align}
Finally, the problem is closed with lateral boundary conditions that define the Sturm\textendash Liouville problems (\ref{eq:Helmholtz_t_laplacian}-\ref{eq:Helmholtz_t_laplacian_z}). Since the upper wall is electrically insulated, we assume that the current injected at the electrode escapes radially at infinity \citep{baker_dimensionality_2015}. Axisymmetry imposes that for $r_e/r_0\gg1$, $j_r\sim r^{-1}$ so the corresponding 
boundary condition for the Sturm\textendash Liouville problem (\ref{eq:Helmholtz_t_laplacian}) is $\lim_{r\rightarrow\infty} j_r=0$.
However, since we solve the problem numerically, the boundary condition at infinity is approximated by one at finite radius $r_e\gg r_0$. Axisymmetric solutions of the Sturm\textendash Liouville problems (\ref{eq:Helmholtz_t_laplacian}) are Bessel functions of the first kind $J_1(\kappa_\perp^{i} r)$, 
where {the transverse wavenumber} $\kappa_\perp^i$ is $J_1$'s $i^{\rm th}$ root scaled by the dimensionless radius of the vessel $r_e$.
\subsection{Boundary conditions and general form of the solution}
With these conditions in the horizontal plane, the general form for $\rmbt{= \rmb_\perp \bm\cdot \unitt}$ and $u_\theta{= \boldsymbol u_\perp \bm\cdot \unitt}$ {is} 
\begin{align} \label{sol:u_b_Bessel_expension}
\{u_\theta,\rmbt\}{(r,z,t)}= \sum_{i=1}^{N_\perp}\{u_\theta^i,\rmbt^i\}{(z,t)}J_1\left(\kappa_\perp^i r\right),
\end{align}
with
\begin{align}\label{eq:general_form_dist}
\{u^i_{\theta}, \rmbt^i\}\left(z,t\right)&= \left\lbrace\mathrm{e}^{s_1^i\,z} \left(\left\lbrace U_1^i,B_1^i\right\rbrace \cos \left(t + \kappa_{1}^i\,z\right) - \left\lbrace U_2^i,B_2^i\right\rbrace \sin\left(t + \kappa_{1}^i\,z\right) \right)\right.\notag\\
&+  \mathrm{e}^{-s_1^i\,z} \left(\left\lbrace U_3^i,B_3^i\right\rbrace \cos \left(t - \kappa_{1}^i\,z\right) - \left\lbrace U_4^i,B_4^i\right\rbrace \sin\left(t - \kappa_{1}^i\,z\right) \right) \notag\\
&+ \mathrm{e}^{s_2^i\,z} \left(\left\lbrace U_5^i,B_5^i\right\rbrace \cos \left(t + \kappa_{2}^i\,z\right) - \left\lbrace U_6^i,B_6^i\right\rbrace \sin\left(t + \kappa_{2}^i\,z\right) \right) \notag \\
&+ \left. \mathrm{e}^{-s_2^i\,z} \left(\left\lbrace U_7^i,B_7^i\right\rbrace \cos \left(t - \kappa_{2}^i\,z\right) - \left\lbrace U_8^i,B_8^i\right\rbrace \sin\left(t - \kappa_{2}^i\,z\right) \right) \right\rbrace,
\end{align}
{and where $\left\{\kappa_1^i,\,\kappa_2^i\right\}$ and $\left\{s_1^i,\,s_2^i\right\}$ are the wavenumbers and spatial attenuations associated with the transverse wavenumber $\kappa_\perp^i$, respectively. 
To obtain the solution, we calculate $\left\{\kappa_1^i,\,\kappa_2^i\right\}$ and $\left\{s_1^i,\,s_2^i\right\}$ with the appropriate dispersion relation: 
for the solution within the propagative low-$\Rm$ approximation, we use Eq. (\ref{eq:disp_relation}), whereas for the solution within the QSMHD approximation, we use Eq. (\ref{eq:disp_relation_QS}). 	
Then, to find the real coefficients $\left\{U_l^i,B_l^i\right\}_{l=1..8}$, we expand the boundary condition (\ref{eq:modelled_magnetic_forcing}) as }
\begin{align}
	b_{w}\left(r,t\right)=\sum_{i=1}^{N_\perp} \left[\dfrac{2}{{\left( r_e\,\mathrm{J_2} \left(\kappa_\perp^i\,r_e\right)\right)}^2}\,\displaystyle{\int_0^{r_e}} \,\zeta\, b_{\theta}^w\left(\zeta\right)\,\mathrm{J_1} \left(\kappa_{\perp}^i\dfrac{\zeta}{r_e}\right)\mathrm{d\zeta}\right] \, J_1(\kappa_\perp^i r) \cos\left({t}\right).
\end{align}
From the above equation, $\bm B_w^i$ and $\bm U_w^i$ are calculated for each transverse wavenumber $\kappa_\perp^i$ so that they express the boundary conditions on the magnetic and velocity perturbations for each term of the Bessel\textendash Fourier series,
\begin{align}\label{coef_BC_on_B_given_kt}
	\bm B_w^i &= \dfrac{2}{{\left( r_e\,\mathrm{J_2} \left(\kappa_\perp^i\,r_e\right)\right)}^2}\,\displaystyle{\int_0^{r_e}} \,\zeta\, b_{\theta}^w\left(\zeta\right)\,\mathrm{J_1} \left(\kappa_{\perp}^i\dfrac{\zeta}{r_e}\right)\mathrm{d\zeta} \,\begin{bmatrix}
		1 & 0 & 0 &	0 \end{bmatrix}^\mathrm{T},
\end{align}
with
\begin{align}\label{coef_BC_on_U_given_kt}
	\bm U_w^i &= \begin{bmatrix}
		0 & 0 & 0 &	0 \end{bmatrix}^\mathrm{T}.
\end{align}
The expressions of $\bm U_w^i$ and $\bm B_w^i$ can be easily adapted to reflect different types of electrical or mechanical forcing. For example, waves can be forced purely mechanically with an oscillatory rotation of electrically insulating top or bottom walls (or both), or electromagnetically as in the present study. 
Former experiments based on the generic geometry of an axisymmetric vessel subject to an axial magnetic field involve more complex types of forcing at the bottom wall: \cite{lundquist_1949} mechanically forced a wave, using a disk with radial strips at the bottom of the vessel. While the actuation was mechanical and modelled as such by Lundquist, it is possible that the fluid between the strips acted as a solid electrical conductor, whose motion induces electromagnetic forcing.  \cite{lehnert_1954} indeed used such indirect magnetic forcing, by driving a copper disk in oscillatory rotation, thereby inducing an axial current. Whether electromagnetic or mechanical, both experiments would be modelled by a different radial distribution of velocity or magnetic field at the boundary than the one in the present study. In any case, the question of which boundary conditions best represent these two experiments remains one to clarify, as significant discrepancies exist between models and experiments {in these papers}. 
 
{Once the boundary conditions are set for each transverse wavenumber (\ref{coef_BC_on_B_given_kt}-\ref{coef_BC_on_U_given_kt}), the unknown coefficients  $\left\{U_l^i,B_l^i\right\}_{l=1..8}$ are determined using the Navier-Stokes equation (\ref{eq:linear_ns_adm}) and the Sturm\textendash Liouville problem (\ref{eq:Helmholtz_t_laplacian}-\ref{eq:Helmholtz_t_laplacian_z}) (the expression of these coefficients is detailed in appendix \ref{App:exp_for_U_B}).	 
}

{Finally}, for the purpose of modelling the present experiment, where the flow is mapped through electric potential measurements at the Hartmann walls, we need theoretical expressions for these quantities. The electric field $\boldsymbol{E}$ and the electric potential gradient $\boldsymbol{\bm\nabla}\phi$ along the transverse directions can be deduced from the magnetic and velocity fields solutions. The radial electric field {$E_r= \bm{E}\bm\cdot \unitr$} is directly obtained from $u_\theta$ and $b_\theta$ using Ohm's law $\boldsymbol j = \boldsymbol{E} + \boldsymbol{u} \times \unitz$ and Amp\`ere's law,
\begin{equation}\label{eq:exp_Er}
	E_r = -\partial_z \rmbt - u_\theta, 
\end{equation}
where the electric field is normalized by $E_0 = \sigma u_0 B_0$. {The radial component of the potential gradient $\partial_r\phi= \bm\nabla \phi \bm\cdot \unitr$ is calculated from $E_r$ and from the magnetic vector potential $\bm A$, where $\bm b= \bm \nabla \times \bm A$. The radial component of $\bm A$ is first obtained using the Coulomb gauge $\bm \nabla \bm\cdot \bm A = 0$:
	\begin{equation}
	 A_r = \bm\Delta^{-1} \partial_z \rmbt,  \label{eq:exp_Ar}
	\end{equation}
where $\bm\Delta^{-1}$ is the inverse of the vector Laplacian operator. Thus, using the Faraday law and Eqs. (\ref{eq:exp_Er}-\ref{eq:exp_Ar}), 
\begin{align} \label{eq:exp_radial_gradPhi}
	\partial_r\phi = -E_r -  \partial_t A_r = u_\theta  + \left(\partial_z - \partial_t\bm\Delta^{-1} \partial_z \right) \rmbt.
\end{align}
Assuming the same form for $E_r$ and $\partial_r\phi$ as for $u_\theta$, 
with corresponding coefficients $E^i_{l=1..8}$ and $\mathit{\Phi}^i_{l=1..8}$, 
the solutions for $E_r$ and $\partial_r\phi$ are determined using (\ref{eq:exp_Er}) and (\ref{eq:exp_radial_gradPhi}), respectively (see appendix \ref{App:exp_for_E_Phi}).
}
\subsection{Quantities for flow diagnosis}\label{sbsec:diagnosis_qties}
\begin{figure}[h]
	\centering
	\includegraphics[width=0.9\linewidth]{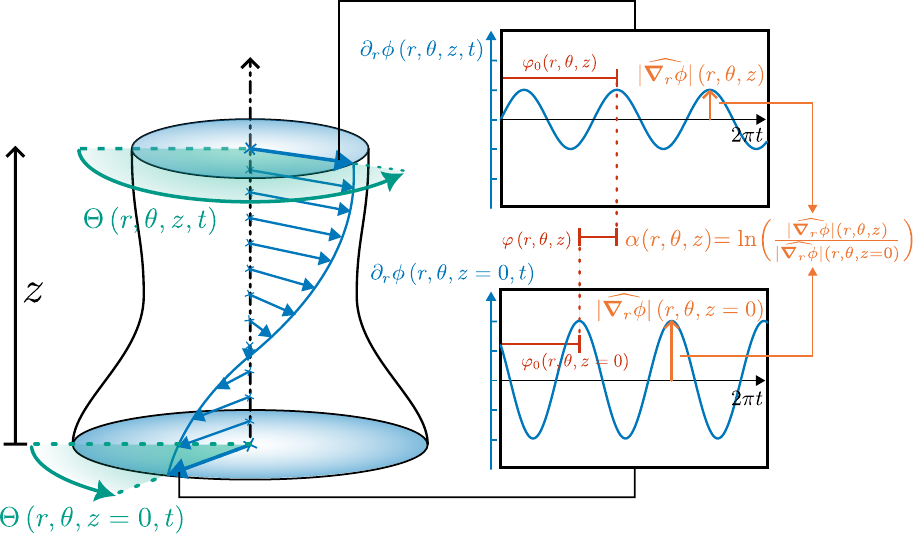}
	\caption{Illustration of the different diagnosis quantities for the oscillating flow: the local phase shift $\varphi\left(r,\theta,z\right) = \varphi_0\left(r,\theta,z\right) - \varphi_0\left(r,\theta,z=0\right)$; the local attenuation  coefficient $\alpha\left(r,\theta,z\right)= \ln \left(|\widehat{\bm\nabla_r\phi}\left(r,\theta,z\right)|/ |\widehat{\bm\nabla_r\phi}\left(r,\theta,z=0\right)\right)|$; the horizontal polar angle $\Theta\left(r,\theta,z,t\right)$ of the horizontal electric potential gradient ${\bm\nabla}\phi$.}
	\label{fig:diagnosis_qty}
\end{figure}

Being now in a position to express solutions semianalytically,  
we define three quantities (represented schematically in figure \ref{fig:diagnosis_qty}) to best characterise how oscillations imposed at the bottom wall propagate, damp and rotate across the layer, in both theory and experiments. 

\begin{enumerate}
\item The local phase shift with respect to the bottom wall
based on {$\partial_r\phi$, the electric potential gradients along $\unitr$ (or based on $\bm \nabla\phi \bm\cdot \unitx= \partial_x\phi $, in Cartesian coordinates, in $\S$ \ref{sec:exp_results}),}
\begin{equation}
	\varphi\left(r,\theta,z\right) = \varphi_0\left(r,\theta,z\right) - \varphi_0\left(r,\theta,z=0\right),
\end{equation}
where {$\varphi_0\left(r,\theta,z\right)= \arg\left\{\int_{0}^{t_s} \unitr\bm\cdot\bm\nabla\phi\left(r,\theta,z,t\right) e^{-j t}\mathrm{d}t \right\}$} is the phase of $\partial_r\phi$, obtained analytically, {and $t_s$ is the duration of a time series}. 

\item The local attenuation coefficient based on $\partial_r\phi$ {(or on $\partial_x\phi $, in $\S$ \ref{sec:exp_results})},
\begin{equation}
	\alpha \left(r,\theta,z\right) = \ln \left(\frac{|\widehat{\bm\nabla_r\phi}|\left(r,\theta,z\right)}{|\widehat{\bm\nabla_r\phi}|\left(r,\theta,z=0\right)}\right),
\end{equation}
where $|\widehat{\bm\nabla_r\phi}|\left(r,\theta,z\right)= {2\left|\int_{0}^{t_s}\unitr\bm\cdot\bm\nabla\phi\left(r,\theta,z,t\right) e^{- j t}\mathrm{d}t \right|}$ is the amplitude of the oscillation of $\partial_r\phi$ at unit angular frequency (dimensionally $\omega$), also calculated analytically.
\item The horizontal polar angle $\Theta$ of ${\bm\nabla}\phi$ with respect to the origin
\begin{equation}
	\Theta \left(r,\theta,z,t\right) = -\arctan \left(\frac{\sin\theta\,\unitr\bm\cdot\bm\nabla\phi\left(r,\theta,z,t\right) +\cos\theta\,\unitt\bm\cdot\bm\nabla\phi\left(r,\theta,z,t\right)}{\cos\theta\,\unitr\bm\cdot\bm\nabla\phi\left(r,\theta,z,t\right) -\sin\theta\,\unitt\bm\cdot\bm\nabla\phi\left(r,\theta,z,t\right)}\right).
\end{equation}
The first two quantities are defined so as to return the spatial attenuation and phase shift of a pure sinusoidal planar wave of the form $\exp\left(\alpha z\right)\sin\left(\kappa z - \omega t\right)$. 
For axisymmetric flows, $\Theta$  reduces to $\theta$ and thus is not of interest but will come in hand in $\S$\ref{sec:exp_approach} and $\S$\ref{sec:exp_results}, where the flow is forced with four electrodes and thus no longer axisymmetric.
\end{enumerate}

\subsection{Numerical solver}
\label{sec:numerical_system}
An in-house MATLAB code was developed to numerically solve the axisymmetric one-electrode problem, i.e. to evaluate time-dependent fields $u_\theta$, $\widetilde{b}_\theta$, $\partial_r \phi$ and $E_r$ on a discrete grid.
The code first solves for $s_i,\kappa_{zi}$ in (\ref{eq:kz_i}) and (\ref{eq:sz_i}) and then solves the $N_\perp$ systems of 16 equations (\ref{eq:BC_forcing}-\ref{eq:passage_U_vers_B}) to determine coefficients $U_l^i$ and $B_l^i$. Because $|s_i|$ and 
$\kappa_{1}$ scale with $\Ha$ and appear as exponential arguments, the variable-precision arithmetic package in MATLAB was used to ensure a sufficient accuracy in computing the solutions, with the number of digits set to 320. The code takes as input $\Rn$, $\Rv$ and $\Ha$, the radial location $r$, the number of transverse modes $N_{\perp}$, the outer radius $r_e$ and the requested time interval for evolution.

The discretisation in time and in the $z$ direction are for display only and do not affect the precision of the result. Similarly the precision of the solution for each mode $\kappa_\perp^i$ depends only on the numerical precision set. On the other hand, the approximation of the boundary condition from a Fourier\textendash Bessel expansion leads to two errors: first, the finite number of modes in the series $N_\perp$ incurs a discretisation error; second, an error on current conservation appears by setting boundary condition $J_r=0$ at a finite radius $r_e$ instead of infinity.
We assess both type of errors relative to a high resolution run at $N_\perp^{\rm max}=2500$ for $r_e=\left\{0.66, 1.32, 2.65\right\}$, respectively, as
\begin{equation}
	\varepsilon \left(r_e,N_{\kappa_\perp}\right)= \left.\frac{\parallel\partial_r\phi\left(N_{\kappa_\perp}\right) - \partial_r\phi \left(N_{\kappa_\perp}^{\rm max}\right)\parallel}{\parallel\partial_r\phi\left(N_{\kappa_\perp}^{\rm max}\right)\parallel}\right.,
\end{equation}
and 
\begin{equation}
\varepsilon_b \left(r_e\right)= \frac{\parallel b_{w}\left(r,t=0,r_e\right) - b^w_\theta\parallel}{\parallel b^w_\theta\parallel},
\end{equation}
where $\parallel \cdot \parallel$ represents the $\mathcal{L}^2$ norm. 
Convergence tests performed with the propagative low-$\Rm$ model show that $\varepsilon_b\sim r_e^{-2}$ for $r_e\gtrsim 0.6$ and 
{ $\varepsilon \leq 10^{-10}$ when $N_\perp \geq \left\{400,\,800,\, 1700 \right\}$ for $r_e= \left\{0.66,\,1.32,\, 2.65\right\}$, respectively.
}	
Based on this result, a good compromise between numerical cost and precision is reached by setting  $r_e$ and $N_\perp$ so as to ensure $\varepsilon_b\leq1\%$ on $b_w$ for any $r$, and $\varepsilon\leq  10^{-10}$ for all forthcoming calculations.
The same test is conducted on the QSMHD model, with similar results, so that forthcoming comparison between the two models is conducted with the same number of modes $N_{\perp}$.
\subsection{Propagative resonance versus oscillating diffusive maxima}
\label{sb_sec:AW_characteristics}
\begin{figure}[h!]
	\begin{subfigure}[c]{0.5\linewidth}
		\centering
		\includegraphics[width=.98\linewidth]{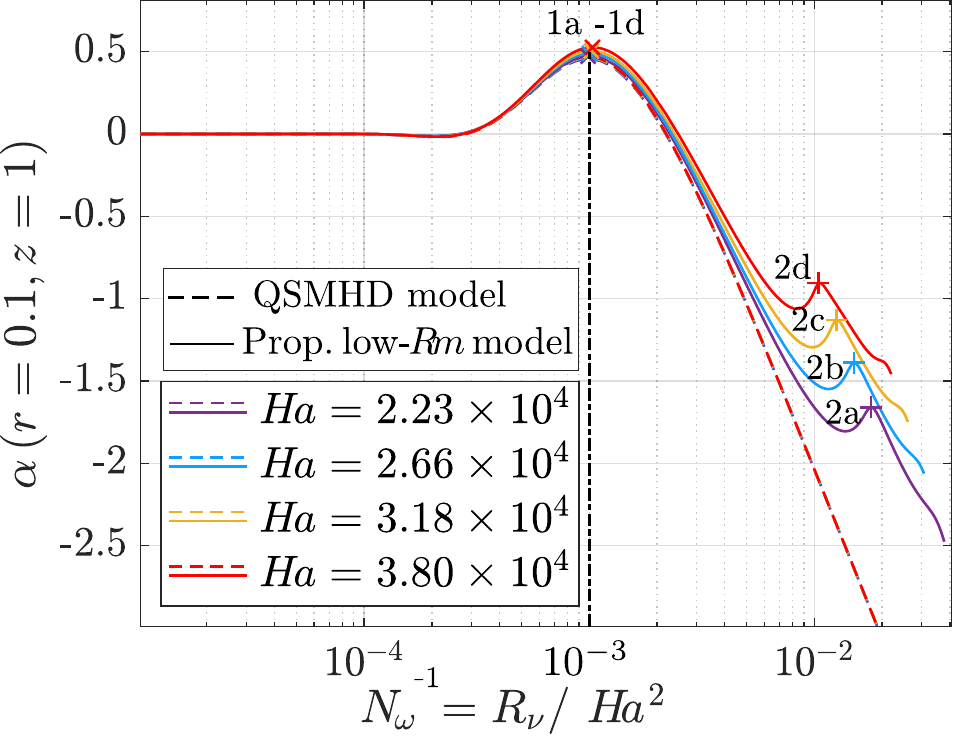}
		\caption{}\label{fig:att_coef_vs_Nomega}
	\end{subfigure}
	\hfill
	\begin{subfigure}[c]{0.5\linewidth}
		\centering
		\includegraphics[width=.99\linewidth]{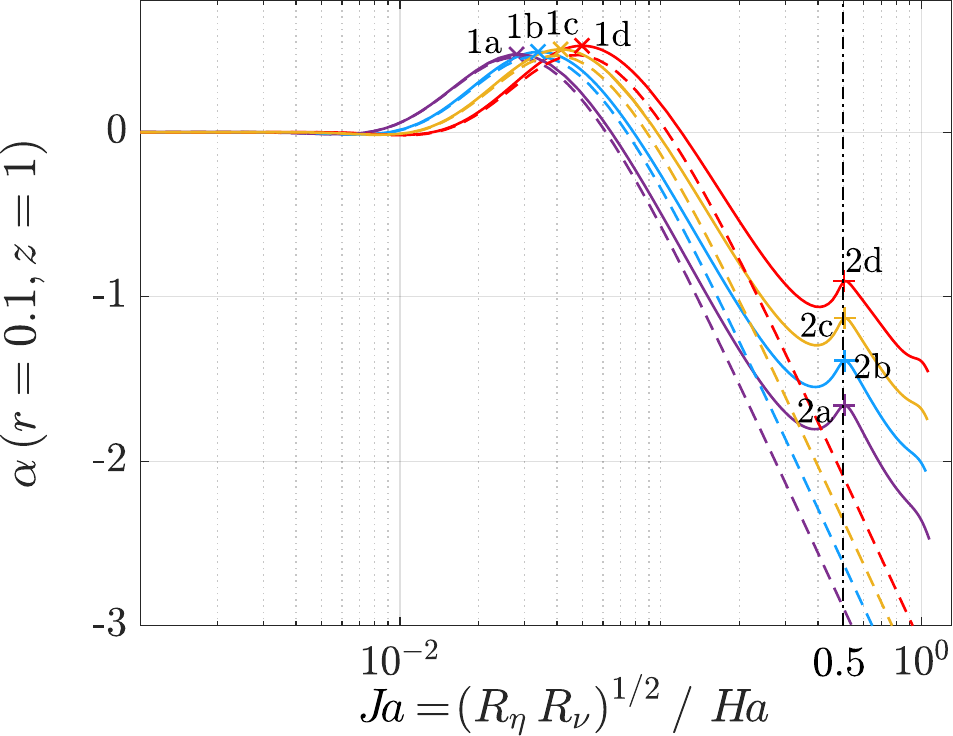}
		\caption{\label{fig:att_coef_vs_LuOmega}}
	\end{subfigure}
	\caption{\label{fig:att_coef_vs_Lu_and_N_omega} Attenuation coefficient $\alpha$ against $\NOmega^{-1}$ (a) and $\Ja$ (b) for $\Ha = \left\{1.9 \times 10^4, 2.66 \times 10^4, 3.18 \times 10^4, 3.8 \times 10^4\right\}$: solid lines (\protect\normalLine), propagative low-$\Rm$ model; dashed lines (\protect\dashedLine), QSMHD model. Calculations are performed at $r=0.1$. The attenuation coefficient $\alpha$ is calculated at $z=1$.}
\end{figure}

We first seek to identify the parameter regime {where propagation occurs and the purely diffusive regime}. To this end, we seek solutions of the one-electrode problem with both the propagative low-$\Rm$ model and the QSMHD model that cannot produce waves for $\Ha\in\left\{2.23, 2.66, 3.18, 3.80\right\}\times 10^4$ and $\Rn\in[4.3 \times {10^{-5}} ,5.1 \times 10^{1}]$. These values are chosen to match the experimentally accessible range. Waves are expected where discrepancies exist between the models. Figure \ref{fig:att_coef_vs_Lu_and_N_omega} represents the variations of the attenuation coefficient $\alpha\left(r=0.1, z=1\right)$ with $\Rn$, renormalised with two different parameters designed to highlight {each of the two regimes}.
\begin{figure}[h!]
	\centering
	\includegraphics[width=1.\linewidth]{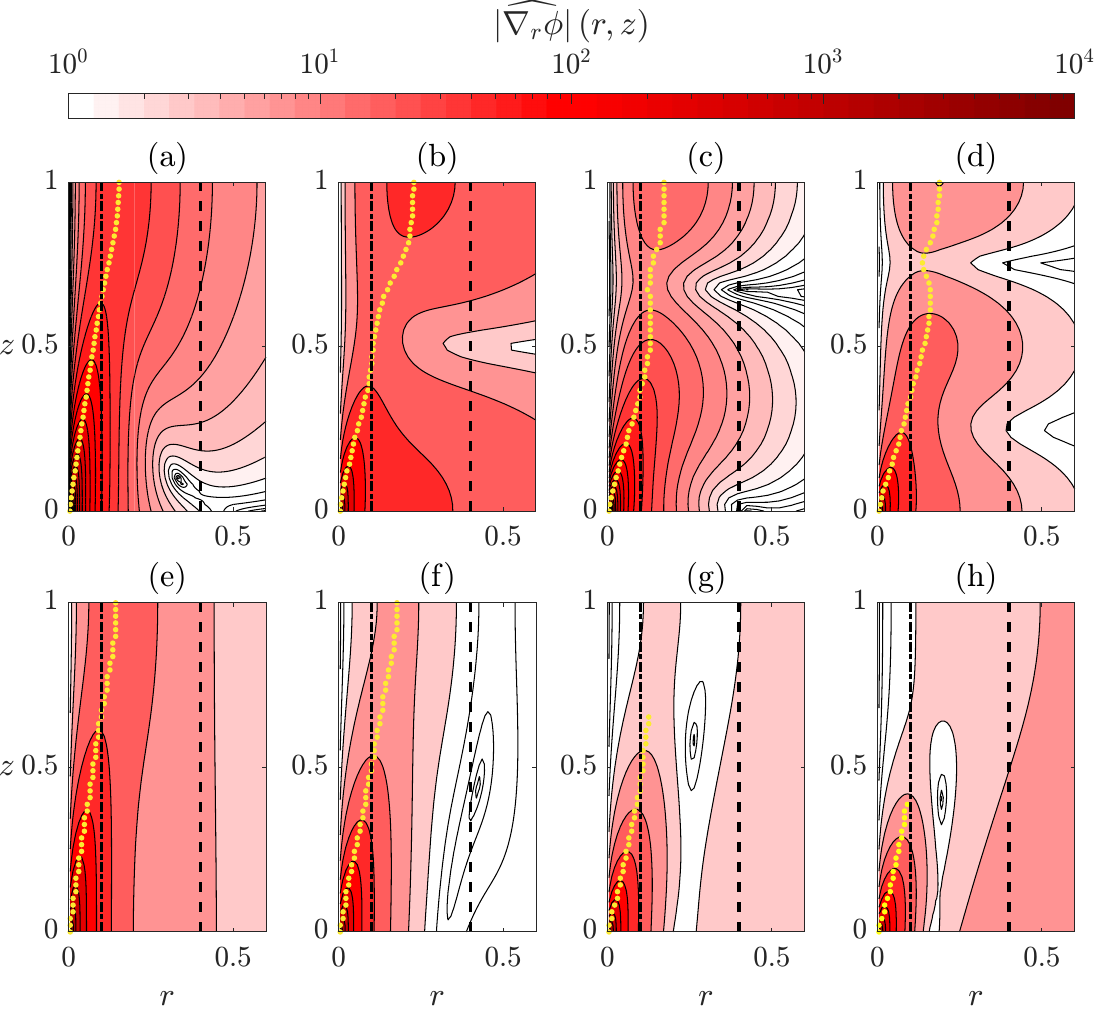}
	\caption{Contours of $|\widehat{\bm\nabla_r\phi}|$ at $\Ha = 3.8 \times 10^4$ and for $\Ja = \{(a,e)\,0.25$;  $(b,f)\,0.5$; $(c,g)\,0.75$; $(d,h)\,1\rbrace$: $(a\textendash d)$ propagative low$-\Rm$ model; ($e\textendash h$) QSMHD model. The vertical dash\textendash dotted line (\protect\dashDottedLine) is located at $r=0.1$ and the vertical dashed line (\protect\dashedLine) at $r=0.4$.\label{fig:mapping_ampl_Ha_Reta}. {The dotted line (\protect\dottedLine) shows the radial location $r_m(z)$ of the maximal velocity.}}
\end{figure} 
The first parameter $\NOmega =\Ha^2/\Rv$ represents the ratio between the Lorentz force and the acceleration term {$\partial_t\bm u$}, present in both models. As such, it does not account for propagative phenomena. We call it the {oscillation parameter} by analogy with the interaction parameter that measures the ratio of Lorentz force to the inertial terms in QSMHD \citep{moreau1990}.
The solutions of the QSMHD and the propagative low-$\Rm$  models collapse for all $\Ha$ numbers in the limit $\NOmega^{-1} \rightarrow 0$ and exhibit a maximum at $\NOmega^{-1} = 1 \times 10^{-3}$. As such, this maximum does not stem from propagative phenomena. The solution of both models starts to diverge around this maximum (marked 1a\textendash 1d on figure \ref{fig:att_coef_vs_Lu_and_N_omega}), with a relative error from $3\%$ to $10\%$ for $2.23 \times 10^4\leq \Ha \leq 3.18 \times 10^4$. The two models diverge significantly more around a second set of maxima (marked 2a\textendash 2d on figure \ref{fig:att_coef_vs_Lu_and_N_omega}), which only exists in the propagative solution. Hence, this second set of maxima necessarily results from a propagative process. 
\begin{figure}[h!]
	\centering
	\includegraphics[width=1.\linewidth]{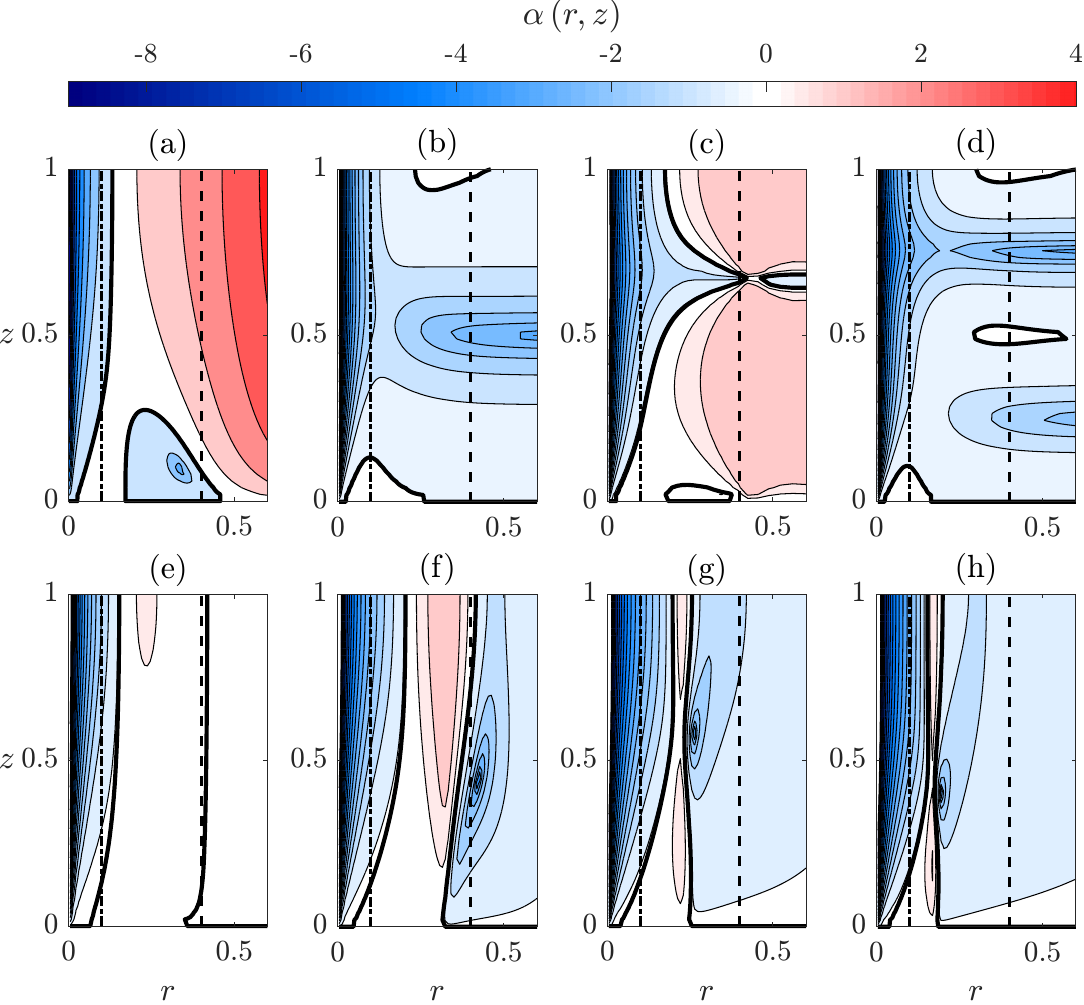}
	\caption{\label{fig:mapping_att_coef_Ha_Reta} Contours of $\alpha(r,z)$ at $\Ha = 3.8 \times 10^4$ and for $\Ja = \{(a,e)\,0.25$;  $(b,f)\,0.5$; $(c,g)\,0.75$; $(d,h)\,1\rbrace$: $(a\textendash d)$ propagative low$-\Rm$ model; ($e\textendash h$) QSMHD model. The vertical dash-dotted line (\protect\dashDottedLine), indicates the radial position $r=0.1$ where values of $\alpha$ are taken for figure \ref{fig:att_coef_vs_Lu_and_N_omega}. The solid line (\protect\thickLine) is the locus of $\alpha(r,z) =0$.
}
\end{figure} 
\begin{figure}[h!]
	\centering
	\includegraphics[width=1.\linewidth]{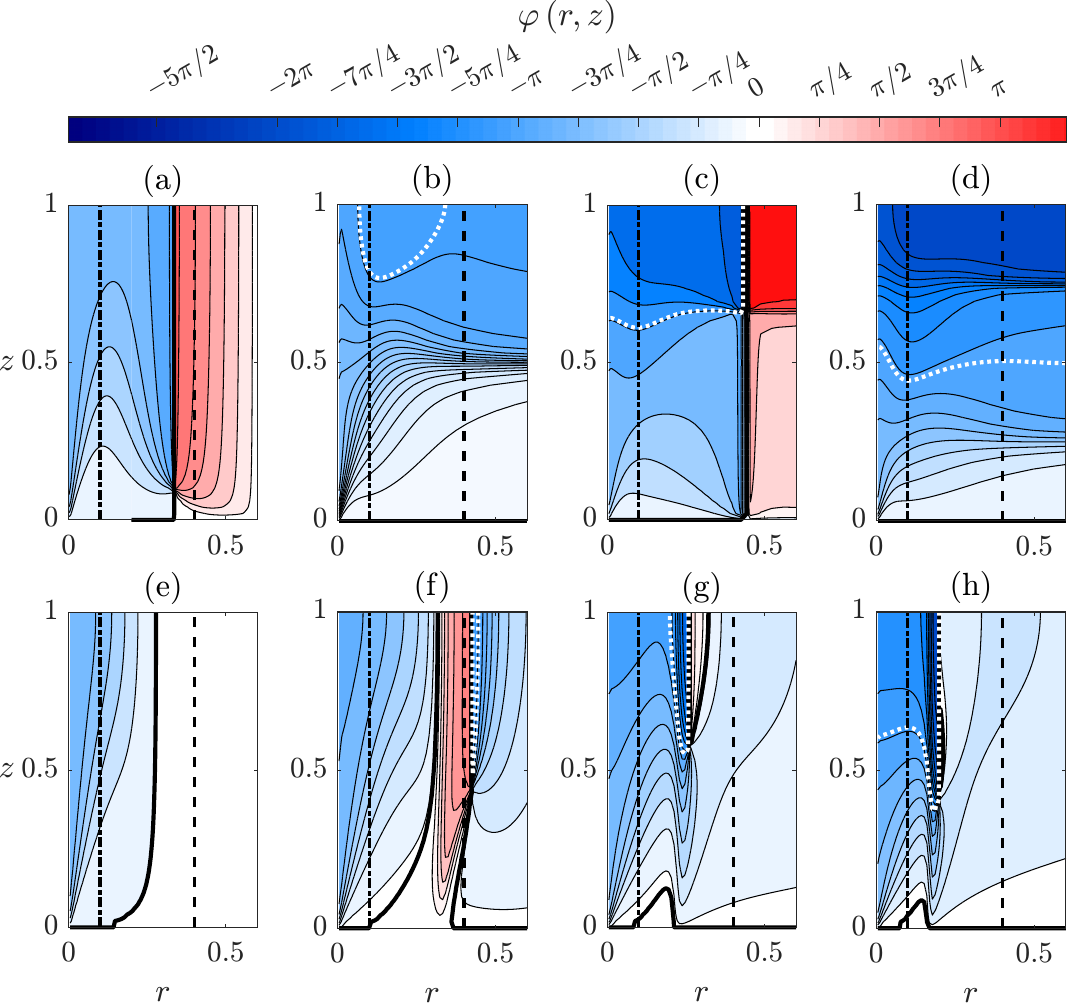}
	\caption{\label{fig:mapping_phase_shift_Ha_Reta} Contours of $\varphi$ at $\Ha = 3.8 \times 10^4$ and for $\Ja = \{(a,e)\,0.25$;  $(b,f)\,0.5$; $(c,g)\,0.75$; $(d,h)\,1\rbrace$: $(a\textendash d)$ propagative low$-\Rm$ model; ($e\textendash h$) QSMHD model.. The solid line (\protect\thickLine) marks the $\varphi =0$ isovalue and the dotted line (\protect\thickDottedLine) the $\varphi =-\pi$ isovalue.}
\end{figure} 
On figure \ref{fig:att_coef_vs_LuOmega}, the attenuation coefficients are plotted against the Jameson number $\Ja= \left(\Rn \Rv\right)^{1/2}/\Ha$. All maxima from the second set (2a-2d) collapse at the same value of $\Ja =0.5$, for all Hartmann numbers. 
Dimensionally, the period associated with these maxima therefore scales with the propagation time of AW across the channel and, thus, suggests a form of {propagative} resonance. 

Both maxima correspond to a reduced attenuation of the oscillations across the channel: however, one is purely diffusive (henceforth referred to as {oscillating diffusive}), and the other one is due to Alfv\'en wave resonances (henceforth {propagative}). 
To understand the precise mechanism behind each of them, we now inspect the spatial distribution of the local amplitude $|\widehat{\bm\nabla_r\phi}| $, the attenuation $\alpha(r,z)$ and the phase $\varphi(r,z)$ of the electric potential gradients.  
Solutions from the QSMHD and propagative low-$\Rm$ models are represented on figures \ref{fig:mapping_ampl_Ha_Reta}, \ref{fig:mapping_att_coef_Ha_Reta} and \ref{fig:mapping_phase_shift_Ha_Reta} for different values of the Jameson numbers $\Ja= $ 0.25 (figures \ref{fig:mapping_ampl_Ha_Reta}a,e, \ref{fig:mapping_att_coef_Ha_Reta}a,e and \ref{fig:mapping_phase_shift_Ha_Reta}a,e), 0.5 (figures \ref{fig:mapping_ampl_Ha_Reta}b,f, \ref{fig:mapping_att_coef_Ha_Reta}b,f and \ref{fig:mapping_phase_shift_Ha_Reta}b,f), 0.75 (figures \ref{fig:mapping_ampl_Ha_Reta}c,g, \ref{fig:mapping_att_coef_Ha_Reta}c,g and \ref{fig:mapping_phase_shift_Ha_Reta}c,g) and 1 (figures \ref{fig:mapping_ampl_Ha_Reta}d,h, \ref{fig:mapping_att_coef_Ha_Reta}d,h and \ref{fig:mapping_phase_shift_Ha_Reta}d,h).
\subsection{Phenomenology of the oscillating diffusive maxima}
\begin{figure}[h!]
	\centering
	\includegraphics[align=b,width=0.75\linewidth]{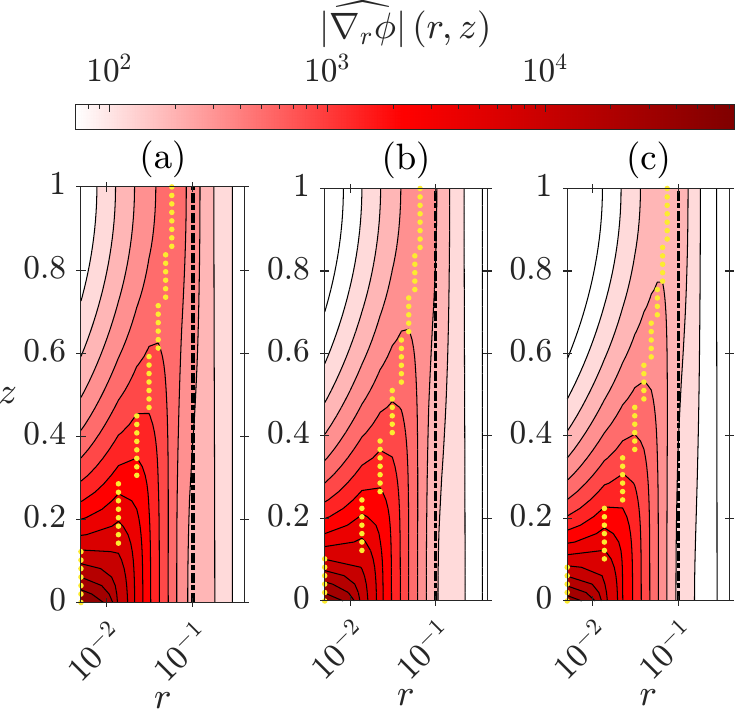}
	\caption{Contours of $|\widehat{\bm\nabla_r\phi}|$ from the QSMHD model at $\Ha = 3.8 \times 10^4$ and for $\NOmega^{-1} = \{(a)\,0.75$; $(b)\, 1$; $(c)\,1.25\}\times 10^{-3}$. {The dotted line (\protect\dottedLine) shows the radial location $r_m(z)$ of the maximal velocity.}\label{fig:ampl_contour_for_diff_Nomega}} 
\end{figure}
Understanding the mechanism underpinning the oscillating diffusive maxima requires a closer look at the region near $r=0$, where velocities are small. Since in the limit $r\rightarrow\infty$, $\vel\rightarrow0$, the flow reaches a maximum velocity
at a radius $r_{\rm m}(z)$ between the close and far regions, {represented by a dotted yellow line on the contour plots of the amplitude of the electric potential gradients in figures \ref{fig:mapping_ampl_Ha_Reta} and \ref{fig:ampl_contour_for_diff_Nomega}.}
The fluid is accelerated in the region $r\leq r_{\rm m}(z)$. In the limit $\Rv\rightarrow0$, this region is governed by the balance between viscous friction in horizontal planes and the Lorentz forces,
\begin{equation}
	\nu \Delta_\perp\sim {\nu}{\Ha^2}\Delta_\perp^{-1}\partial_{zz}^2,
\end{equation}
which, for a lengthscale along $z$ scaling with z, leads to the classical scaling for the thickness of vortex cores in high magnetic fields: $\tilde r_{\rm m}^{(\nu)}/h\sim \Ha^{-1/2} (\tilde z/h)^{1/2}$. Crucially, the vortex core becomes thicker as the axial distance to the injection point increases \citep{sommeria1988_jfm,potherat2000_jfm}.

For $\Rv$ sufficiently large, the {oscillating}, acceleration term $\partial_t\bm u$ overcomes viscous 
friction to balance the Lorentz force and so $\tilde r_{\rm m}^{(\omega)}/h\sim (\Ha^2/\Rv)^{-1/2} \tilde z/h=\NOmega^{-1/2} \tilde z/h$. The transition between these two
regimes takes place when $\tilde r_{\rm m}^{(\omega)}\sim \tilde r_{\rm m}^{(\nu)}$, i.e. when $\Rv\sim\Ha$ at $\tilde z=h$. Since the oscillating diffusive maximum is located at {$\Rv/\Ha = \Ha \NOmega^{-1}\gtrsim 10 \gg 1$ for all values  of $\Ha$ considered, it is well within the regime dominated by the acceleration term.
{The contour plots of $|\widehat{\bm\nabla_r\phi}|$ in figure \ref{fig:ampl_contour_for_diff_Nomega} obtained with the QSMHD model for $\NOmega^{-1} = \left\{0.75,\,1,\,1.25\right\} \times 10^{-3}$, illustrate the oscillating diffusive process at low $\NOmega^{\,-1}$.}
In this regime too, the vortex core becomes wider at a larger distance from the electrode, {as evidenced by the slanted contours of $|\widehat{\nabla_r\phi}|$ in figures \ref{fig:mapping_ampl_Ha_Reta} and \ref{fig:ampl_contour_for_diff_Nomega}}. However, when $z$ reaches the axial wavelength of the oscillations $2\pi/\kappa$, the second derivative $\partial_{zz}$ saturates (at $(\tilde \kappa)^2/(4\pi^2)$) and so does 
the radius of the vortex core. This $z$ variation implies that {for $\tilde r>\tilde r_{\rm m}$, the point $r=\tilde r/h=0.1$ on the bottom wall is always in a region of lower amplitude oscillations than the point $r=0.1$ on the top wall. This can be seen as the $r=0.1$ line crosses an isovalue of $|\widehat{\nabla_r\phi}|$ in all examples at low $N_\omega^{-1}$}. 
{$\tilde r_m^{(\omega)}$ also depends on the forcing frequency. Here $\tilde r_m^{(\omega)}$ becomes larger as $\Rv$ increases, \ie as $\NOmega^{-1}$ increases. This is confirmed in figure \ref{fig:ampl_contour_for_diff_Nomega} where the yellow dotted line which highlights $r_m^{(\omega)}=\tilde r_m^{(\omega)}/h$ along the vessel moves radially outwards as $\NOmega^{-1}$ increases. 
The two different dependences of $\tilde r_m^{(\omega)}$ in the near- and far-region of the electrode incur a maximum in the ratio of amplitudes of the oscillations at the top and bottom walls at a given $\NOmega$ value, $\NOmega^{-1} = 1\times 10^{-3}$ in the present study (illustrated in figure \ref{fig:ampl_contour_for_diff_Nomega}b). This translates into the maximum observed when varying $N_\omega$ whilst measuring the amplitude ratio at a fixed location.}
\subsection{Phenomenology of the propagative maxima}\label{sb_sec:th_propagative_mech}
\begin{figure}[h]
	\centering
	\includegraphics[width=0.7\linewidth]{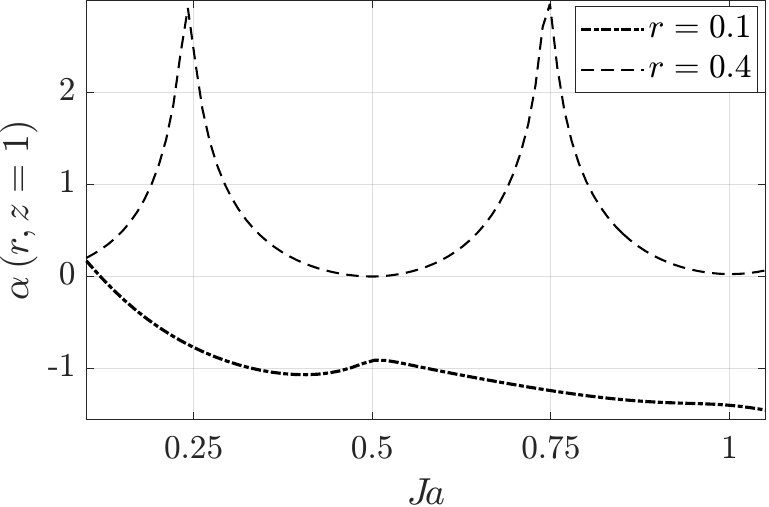}
	\caption{ Here $\alpha$ from the propagative low-$\Rm$ model against $\Ja$ for $r=\left\{0.1, 0.4\right\}$ at $\Ha= 3.8 \times 10^4$.\label{fig:alpha_vs_Luw_diff_r}}
\end{figure} 

The propagative resonance is governed by the reflection of AW on the Hartmann walls, 
as each propagative resonance corresponds to values of $\Rn$ (or of the frequency) for which the wavelength $l_A=V_A/2\pi\omega$ is an integer fraction of the channel height $h$ or, equivalently, $\Ja=\{1/4,1/2,3/4,1...\}$.\\
Far from the electrode, the radial velocity and potential gradients are small compared with the axial ones. This is made visible through the horizontal orientation of $\varphi$ isolines as $r$ increases (figure \ref{fig:mapping_phase_shift_Ha_Reta}). In areas of low radial gradient, such as near $r=0.4$, propagative resonances appear at $\Ja=\left\{1/4, 3/4\right\}$
{i.e.}, respectively, $l_A=\left\{h/4, 3h/4 \right\}$. 
This is illustrated on figure \ref{fig:alpha_vs_Luw_diff_r}, which shows the variations of $\alpha$ with $\Ja$ for two radial locations $r=\left\{0.1, 0.4\right\}$ at $\Ha= 3.8 \times 10^4$.
The maximum in $\alpha$ results from the coexistence of a minimum of $|\widehat{\bm\nabla_r\phi}|$ at the bottom wall and a maximum at the top wall. 
This is also visible from the location of the node and antinode of 
$|\widehat{\bm\nabla_r\phi}|$, $\alpha$ and $\varphi$ along the dashed line (\protect\dashedLine) 
$r=0.4$ {in figures \ref{fig:mapping_ampl_Ha_Reta}$(b,c)$, \ref{fig:mapping_att_coef_Ha_Reta}$(b,c)$ and \ref{fig:mapping_phase_shift_Ha_Reta}$(b,c)$, respectively.  
This case corresponds to theory and experiments by} \cite{jameson_1964}, whose resonances also appear at $\Ja=1/4$ and $3/4$, in our notations.
\begin{figure}
\centering
\includegraphics[width=0.7\linewidth]{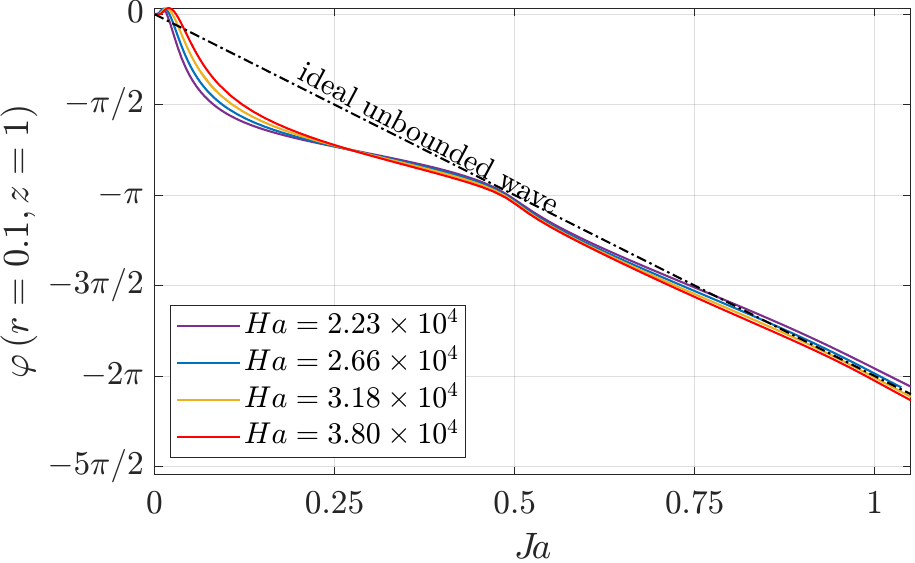}
\caption{Phase shift $\varphi$ from the propagative low-$\Rm$ model against $\Ja$ at $r=0.1$ for $\Ha = \left\{1.9 \times 10^4, 2.66 \times 10^4, 3.18 \times 10^4, 3.8 \times 10^4\right\}$. The dash\textendash dotted line (\protect\dashDottedLine) $\varphi$ is for an ideal unbounded wave. \label{fig:phase_shift_vs_S_omega}}
\end{figure} 
Close to the electrode, by contrast, radial gradients increase significantly. Comparing the values of $\alpha$ at $r=0.1$ with those at $r=0.4$ in figure \ref{fig:mapping_att_coef_Ha_Reta} shows that the attenuation coefficient decreases in absolute value in this region. Hence, waves are more strongly damped in areas of higher transverse gradients, because of the extra Joule dissipation that these incur.
In this region, \eg at $r=0.1$, the propagative resonance appears at $\Ja= 0.5$ (marked 2a\textendash 2d on figure \ref{fig:att_coef_vs_LuOmega}). Plots of $\varphi$ against $\Ja$ at $r=0.1$ for different values of $\Ha$ on figure \ref{fig:phase_shift_vs_S_omega} reveal that the phase shift between the two 
Hartmann walls is $-\pi$ at this resonance. Thus, propagative resonances in strong radial gradients take place in phase opposition, which is confirmed by the two antinodes of $|\widehat{\bm\nabla_r\phi}|$ located at the Hartmann walls for this $\Ja$ value (see figure \ref{fig:mapping_ampl_Ha_Reta}b). 

Lastly, an ideal AW propagating between the Hartmann walls would incur a phase shift with a linear dependence on $\Ja$. From figure \ref{fig:phase_shift_vs_S_omega}, by contrast, $\varphi$ exhibits a nonlinear dependence on $\Ja$. The discrepancy to linearity is 
particularly pronounced when $\Ja\lesssim0.8$. This regime overlaps the regime of wave propagation for $\Ja\gtrsim0.1$, so in this regime, dissipation causes AW to become dispersive. This effect, however, progressively vanishes at higher values of $\Ja$. 
Hence, in this limit, a nearly  {dispersionless} propagative regime is recovered, despite the dissipation.

\section{Experimental methods}\label{sec:exp_approach}
\subsection{The FlowCube experimental device}
\label{sec:flowcube}
The experimental device is an upgraded version of the FlowCube facility \citep{klein_potherat_2010,pk2014_jfm, baker_controlling_2017}. It consists of a modular cuboid vessel of inner height $h= 10$ cm and width $L = 15$ cm. The sidewalls are electrically insulated and made of polycarbonate. The top and bottom Hartmann walls in contact with the fluid are printed circuit boards made of an FR4 epoxy layer and a ROGER 4003 ceramic layer acting as an insulator, all mounted on polyamide-coated brass frame. Both are fitted with injection electrodes to drive the flow, and probes to measure electric potentials. The working liquid metal, hermetically enclosed inside the vessel, is {a eutectic} alloy of gallium, indium and tin that is liquid at room {temperature}, of electric conductivity $\sigma = 3.4\times 10^{6} \,\mathrm{S/m} $,  density $\rho = 6400\,\mathrm{kg\,m^{-3}}$, kinematic viscosity $\nu = 3.7\times 10^{-7}\mathrm{m^2\,s^{-1}}$ ({i.e.} the magnetic Prandtl number is $\Pm= \Rn/\Rv= \nu/\eta= 1.6 \times 10^{-6}$). The procedures described in \cite{baker_controlling_2017} ensure hermetic filling {of the vessel and} good electrical contact between the metal and both sets of probes and electrodes.
FlowCube is subjected to a static and uniform axial magnetic field $B_0 \, \boldsymbol{e_z}$ provided by the 12 MW M10 resistive electromagnet of $376$ mm diameter bore at the LNCMI (CNRS) in Grenoble. The field homogeneity over the vessel is approximately $5\%$. The range of magnetic fields used $B_0 =  \left\{ 5.87, 7, 8.37, 10 \right\}\,$T corresponds to Hartmann numbers $\Ha = \left\{1.9 \times 10^4, 2.66 \times 10^4, 3.18 \times 10^4, 3.8 \times 10^4\right\}$.

\subsection{Flow forcing mechanism}

\begin{figure}
	\centering
	\includegraphics[width=0.65\linewidth]{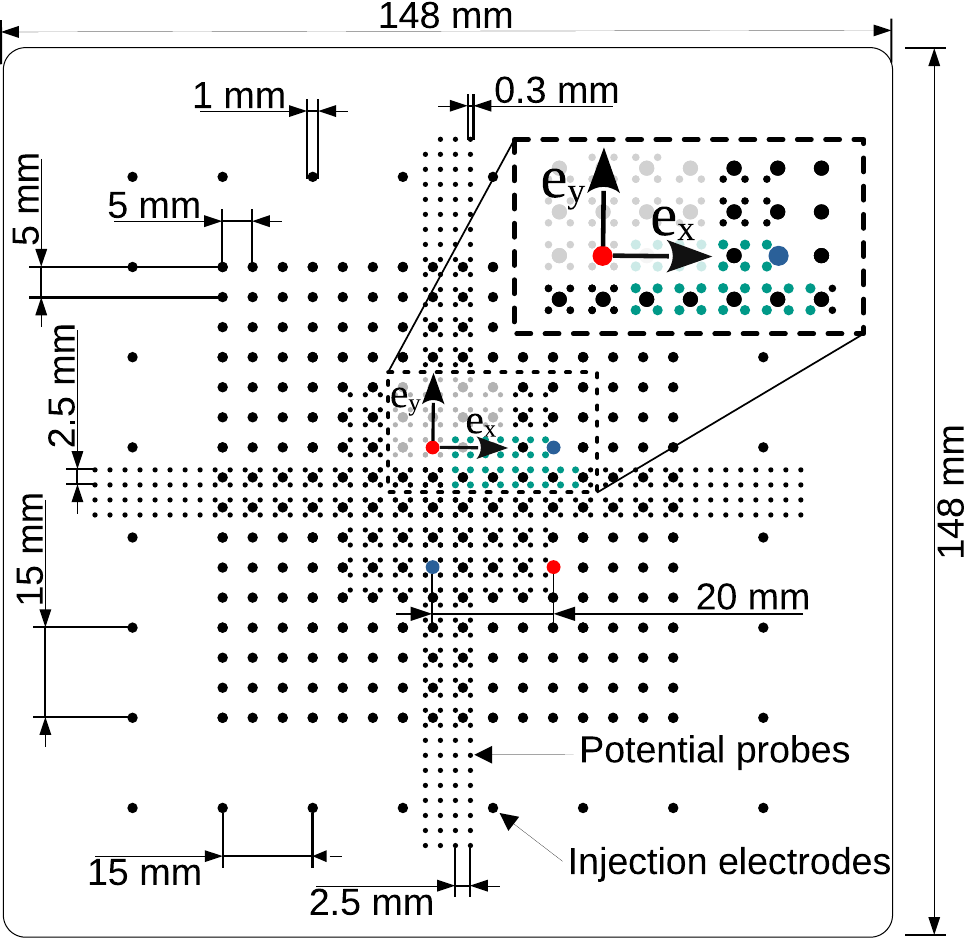}
	\caption{Sketch of the face of the bottom Hartmann wall in contact with the {GaInSn} alloy The injection pattern used during the experiments is represented coloured circles. The red circles are connected to the phase of the AC power supply and the blue ones to the neutral. The blue and red electrodes are in phase opposition. Tail circles represent active potential probes.}
	\label{fig:schema_Hartmann}
\end{figure}

As in \cite{sommeria_experimental_1986}, \cite{klein_experiment_2009} and \cite{baker_controlling_2017}, the flow is driven by injecting a prescribed current through an array of electrodes fitted flush at the bottom Hartmann wall located at $z = 0$. These are arranged on a 16x16 square lattice spaced $5$ mm apart and a ring lattice of height electrodes per side spaced $15$ mm apart (figure \ref{fig:schema_Hartmann}). Each of the electrodes is made of 1 mm diameter copper wire and surface-coated with a thin layer of gold to warrant the electrical contact with the liquid metal. An AC electric current is injected through a subset of these electrodes. In this paper, the corresponding forcing pattern is a near-central square of $ 2 \times 2$ electrodes spaced by $\tilde{d}_0 = 0.02\,$m (non-dimensionally, ${d_0}=\tilde{d}_0/{h} =0.2$), with adjacent electrodes of opposite polarities (figure \ref{fig:schema_Hartmann}). 
The injected current is generated by a HP 33120A function generator feeding into two superimposed KEPCO 400 W BOP 20-20M amplifiers.  
It is sinusoidal, with a frequency range $\omega/\left(2\pi\right) \in \left[1,1330\right]\,$Hz corresponding to $\Rn \in \left[ 4.4 \times 10^{-2},  5.8 \times 10^1 \right]$. 
Different root mean square (r.m.s.) current intensities $|I_{0}|$ per electrode were investigated, from $0.1\,$A to $6.5\,$A. 
To overcome current variations incurred by fluctuations in the contact resistance between metal and electrodes (of the order of $1\times 10^{-2}\Omega$), a constant ohmic resistance of
$2 \Omega \pm 0.25\%$ is added in series to each electrode \citep{pk2014_jfm}. 
Injecting AC current in this way drives an oscillatory flow of controlled frequency, amplitude, and transverse lengthscale given by the pitch $\tilde{d}_0$ between electrodes.
\subsection{Measurement of electric potentials}
The flow is diagnosed using two arrays of electric potential probes fitted flush at the surface of each Hartmann wall. They are arranged on a $2.5$ mm lattice made of a central square of 14$\times$14 probes, and two perpendicular arrays of 4$\times$48 probes spanning the $x$ and $y$ directions of the Hartmann walls  (figure \ref{fig:schema_Hartmann}). The arrays of both walls are aligned opposite each other along $\boldsymbol e_z$.
Additionally, one of four potential probes located at the outer ring of the wall serves as reference for potential measurements. Each potential probe is made of $0.2\,$mm diameter copper wire touching the metal at the wall surface.
From these potential probes, the electric signals are gathered through printed circuit boards embedded into the Hartmann walls and routed externally to two National Instrument PXIe-4303 data acquisition modules housed in a PXI express chassis clocked at $10\,$MHz. 
Each module synchronously records the signals of 32 probes to 24-bit precision at a sampling frequency {$f_s$} of up to 52.2 kHz per channel, with a low r.m.s. noise ($3.2\,\mathrm{\mu\! V}$). 
{Such high acquisition frequencies are required to accurately capture the different physical processes at play in the flow. The diffusive and propagative regimes of the Lorentz force act on respective time scales $\tau_{2D}= (\rho/(\sigma B_0^2))(h/d_0)^{2}$ and $\tau_{A} = h/V_A$, so $f_s \gg \max\left\{\tau_{2D}^{-1},\tau_{A}^{-1} \right\}$ ensures that the both regimes can be captured. For instance, $\tau_{2D}^{-1} \approx 2$ kHz and  $\tau_A^{-1} \approx 1$kHz at the highest magnetic field of $B_0 ={10}$ {T}, where both time scales are are smallest.}
\begin{table}
	\centering
	\begin{tabular}{r l c c c c}
		$B_0$ [T] & & 5.87 & 7 & 8.37 & 10 \\
		$\omega/2\pi$ [Hz]& $\min$ & 1 & 1 & 1 & 1  \\
		&$\max$ & 781 & 933 & 1115 & 1330  \\
		$I_0$ {[A]} & & 1.40 & 1.40 & 1.40 & 1.40 \\
		\hline
		$\Ha$ & & $2.23 \times 10^4$ & $2.66 \times 10^4$ & $3.18 \times 10^4$ & $3.80 \times 10^4$ \\
		$\Lu$ & & $2.8\times 10^1$ & $3.3\times 10^1$ & $4.0\times 10^1$ & $4.8\times 10^1$ \\
		$\Rn$ & $\min$ & $4.36 \times 10^{-2}$ & $4.36 \times 10^{-2}$ & $4.36 \times 10^{-2}$ & $4.36 \times 10^{-2}$  \\
		 & $\max$ & $3.41 \times 10^{1}$ & $4.07 \times 10^{1}$ & $4.85\times 10^{1}$ & $5.79\times 10^{1}$  \\
		${\Rv}$ & $\min$ & $2.69\times 10^{5}$ & $2.69\times 10^{5}$ & $2.69\times 10^{5}$ & $2.69\times 10^{5}$  \\
		 & $\max$ & $2.10\times 10^{8}$ & $2.50\times 10^{8}$ & $2.99\times 10^{8}$ & $3.57\times 10^{8}$  \\
		${\Rey}_{0}$ & & $1.33\times 10^{4}$ & $1.33\times 10^{4}$ & $1.33\times 10^{4}$ & $1.33\times 10^{4}$ \\
		\hline
		${\Rey}^{\mathrm{bot}}_{u}$&$\min$ & $6.25\times 10^{1}$ & $5.50 \times 10^{1}$ & $5.35 \times 10^{1}$ & $5.13 \times 10^{1}$  \\
		 & $\max$ & $2.96 \times 10^{3}$ & $4.90 \times 10^{3}$ & $5.94\times 10^{3}$ & $6.37\times 10^{3}$  \\
		${\Rey}^{\mathrm{top}}_{u}$ & $\min$ & $1.79 $ & $1.72$ & $2.41$ & $3.95$  \\
		 & $\max$ & $2.17 \times 10^{3}$ & $3.88 \times 10^{3}$ & $5.22 \times 10^{3}$ & $5.50\times 10^{3}$  \\
		${\Rm}^{\mathrm{bot}}_{u}$ & $\min$ & $1.02\times 10^{-4}$ & $8.94 \times 10^{-5}$ & $8.69 \times 10^{-5}$ & $8.33 \times 10^{-5}$  \\
		 & $\max$ & $4.80 \times 10^{-3}$ & $8.00 \times 10^{-3}$ & $9.70\times 10^{-3}$ & $1.03\times 10^{-2}$  \\
		${\Rm}^{\mathrm{top}}_{u}$ &$\min$ & $2.91 \times 10^{-6}$ & $2.79 \times 10^{-6}$ & $3.92 \times 10^{-6}$ & $6.42 \times 10^{-6}$  \\
		 & $\max$ & $3.50 \times 10^{-3}$ & $6.30 \times 10^{-3}$ & $8.50 \times 10^{-3}$ & $8.90\times 10^{-3}$  \\
		\end{tabular}
	\caption{\label{table_range_parameter}Range of control parameters and dimensionless numbers investigated in $\S$$\S$ \ref{sb_sec:two_regimes_exp_and_th}\textendash\ref{sb_sec:horizontal_gradient_exp}. {The table also shows the extrema values at the top and bottom walls for the measured Reynolds and magnetic Reynolds numbers.}}
\end{table}
{\subsection{Measured Reynolds and magnetic Reynolds numbers}\label{sb_sec:measured_Re_Rm}
The theory developed in $\S$ \ref{sec:theory_1elec} applies in the limit $\Rey\rightarrow0$ and $\Rm\rightarrow0$. Hence, we need to appraise the extent to which these assumptions are valid in the experimental regimes. 
The flow intensity depends \emph{a priori} on $I_0$ and $\omega/\left(2\pi\right)$. In the QS limit, $\Rn\rightarrow0$. In this limit, and for $\Ha \gg 1$, $\Ha^2/\Rey \gg 1$, Eq. \eqref{eq:u0} is a good estimate for the actual flow velocity. As $\Rn$ increases, however, the high frequency shift in the forcing current prevents velocities from ever reaching this value. For this reason, $u_0$ and the control parameter $\Rey_0$ significantly overestimate the actual velocities and Reynolds numbers.}

{Electric potential velocimetry offers a way to estimate at least the order of magnitude of the actual velocity: with this method, transverse potential gradients obtained from the electric potential signals provide an indirect measurement of the transverse velocity field for MHD experiments \citep{sommeria_experimental_1986,alboussiere_1999,klein_potherat_2010,baker_controlling_2017}. 
Electric potential velocimetry relies on the property that for $\Ha\gg1$ and $N\gg1$, most of the current concentrates in the Hartmann layers, so in the bulk near these layers, $\boldsymbol J/\sigma=-\bm\nabla\phi +\up\times B_0\unitz\simeq0$ (dimensionally). Since these layers are very thin, $\phi$ has essentially the same value as its value at the wall $\phi_W$, so the velocity in the bulk can be obtained by measuring the electric potential gradient at the wall by  $\up\simeq B_0^{-1}\unitz\times\bm\nabla\phi_W$, to a precision $O (\Ha,N^{-1})$ \citep{kljukin_direct_1998}. However, while the $\partial_t \boldsymbol A$ in Ohm's law is neglected in the limit of small frequencies, it becomes important in the experiments we consider here. Not accounting for it in the EPV method may incur an error of up to $45\%$ for the highest frequencies we consider. Nevertheless, even with such an error, the values obtained by classical EPV return a reliable order of magnitude of the actual velocity for the purpose of estimating $\Rey$ and $\Rm$.
Hence, based on the EPV methodology, we build an estimate for the velocity from measurements using
\begin{equation}
U_\perp(z) = B_0^{-1} \left\langle \overline{\big(\left\|\bm\nabla_\perp\phi\right\| -\overline{\left\|\bm\nabla_\perp\phi\right\|} \big)^2}^{1/2}  \right\rangle_{x,y},
\end{equation}
where $\left\langle\cdot\right\rangle_{x,y}$ denotes the operator for spatial averaging in the $\left(x,y\right)$ plane, the overbar stands for time averaging and $\|\bm\nabla_\perp\phi\| = {[{(\bm\nabla\phi\bm\cdot\unitx)}^2 + {(\bm\nabla\phi\bm\cdot\unity)}^2]}^{1/2}$ is the time value of the norm of the transverse potential gradient. From this velocity, we further define the Reynolds numbers and a magnetic Reynolds numbers at the top and bottom walls based on the flow intensity as $\Rey_u^{\rm top/bot}= U_\perp^{\rm top/bot} \tilde d_0/\nu$ and $\Rm^{\rm top/bot}= U_\perp^{\rm top/bot} \tilde d_0/\eta$, respectively.
The typical values of these parameters are reported in table \ref{table_range_parameter}. With $10^{-6}\leq\Rm\leq10^{-2}$, the low-$\Rm$ approximation is indeed better satisfied than in most low-$\Rm$ laboratory experiments where $10^{-3}\leq Rm\leq10^{-1}$. 
Reynolds numbers are also sufficiently low to ensure that nonlinear effects remain extremely small (see other low-$\Rm$ studies for comparison \cite{zikanov2014_amm} and \cite{cassels2019_jfm}).} 
\subsection{Experimental procedure}
The FlowCube offers four dimensional control parameters: the injection scale $\tilde{d_0}$, the r.m.s. value of the electric current injected per electrode $I_{0}$; the magnitude of the magnetic field $B_0$; the frequency of the injected current $\omega/\left(2\pi\right)$. In this study, only the last three parameters were varied so as to independently control non-dimensional parameters $\Rey_0$, $\Ha$ and $\Rn$ ($\Pm=\Rn/\Rv = 1.6\times 10^{-6}$ being set by fluid properties). While variations of {$\Rn$, $\Ha$ and corresponding $\Lu$ (all summarised in table \ref{table_range_parameter})} enable us to test the  linear theory in $\S$ \ref{sec:theory_1elec}, $I_0$ and so $\Rey_0$ were varied for the specific purpose of assessing the linearity of the wave forcing, and thereby the limits of the linear theory. 

Measurements are performed at a constant magnetic field $B_0$, i.e. at constant $\Ha$. Starting from a flow at rest, the electric potentials are first recorded over $30$s. Then the AC power supply is switched on at a set current angular frequency $\omega$ and at a r.m.s. value per electrode of the current $I_{0}$.  
Then, electric potentials are recorded for $30$s, to ensure a relative convergence error on $\phi$ lower than $10^{-2}$ for all cases. 
The current supply is then switched off. The last three steps are repeated for different forcing frequencies to span different values of $\Rn$ and the whole procedure is repeated for different values of the magnetic field to span different values of $\Ha$. 
\section{Experimental results and four-electrode theoretical model}
\label{sec:exp_results}
\subsection{Methodology for the comparison between theory and experiment}
\noindent
To enable direct comparison between experimental data and the prediction of the {propagative} low-$\Rm$ model, the latter is extended to the case of a four-electrodes configuration representing the experimental current injection pattern. Since the model is linear, this is done by translating the solution for one forcing electrode to the location of each of the four forcing electrodes and superimposing the four solutions so obtained, weighing them by the sign of the current they carry. The solution is then expressed in a common Cartesian frame whose origin is at the centre of the top left electrode of the injection pattern on figure \ref{fig:schema_Hartmann}. In this frame, we use two attenuation coefficients and phase shifts defined in a similar way as in $\S$ \ref{sbsec:diagnosis_qties}: 
$\alpha_x$ and $\varphi_{ x}$ are based on $\bm\nabla\phi\bm\cdot\unitx$ while $\alpha$ and $\varphi$ are based on {$\|\bm\nabla_\perp\phi\|$}. Gradients {of electric potential} are evaluated by second-order central finite differences using signals from adjacent probes, so $\alpha_x$ and $\varphi_{x}$ require the signals of two probes while $\alpha$ and $\varphi$ require four.
\subsection{Oscillating diffusive regime versus propagative regime
\label{sb_sec:two_regimes_exp_and_th}}
\begin{figure}[h!]
	\begin{subfigure}[c]{0.5\linewidth}
		\centering
		\includegraphics[width=.97\linewidth]{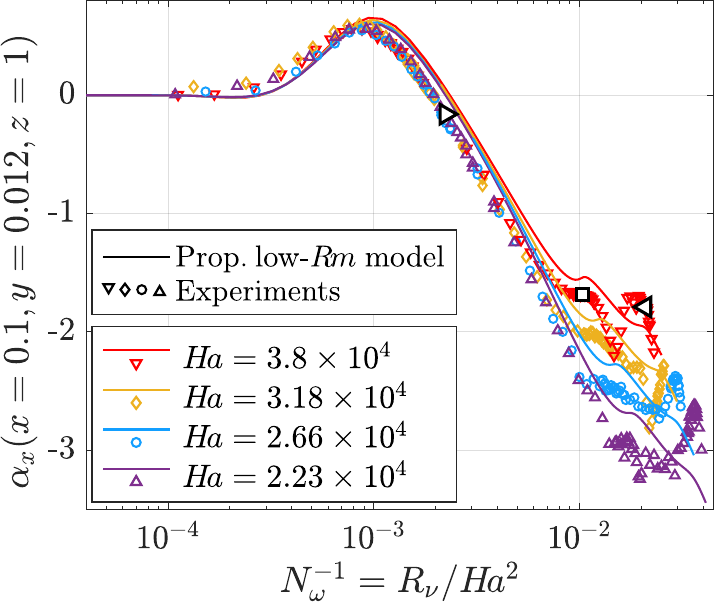}
	\caption{}\label{fig:att_coef_vs_Nomega_exp_th}
\end{subfigure}
\hfill
\begin{subfigure}[c]{0.5\linewidth}
	\centering
	\includegraphics[width=.97\linewidth]{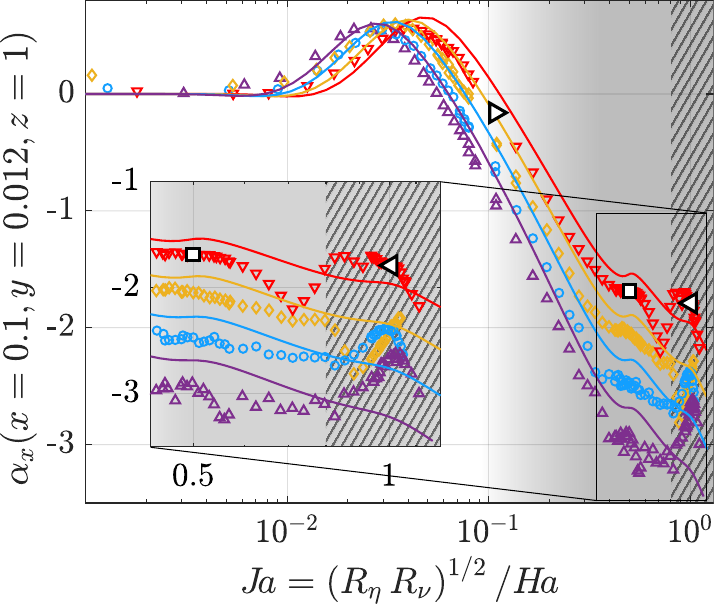}
	\caption{\label{fig:att_coef_vs_LuOmega_exp_th}}
\end{subfigure}
	\caption{\label{fig:att_coef_vs_Lu_and_N_omega_exp_th} Attenuation coefficient $\alpha_{x}$ versus $\NOmega^{-1}$ (a) and $\Ja$ (b) for $\Ha = \left\{1.9\times 10^{4}, 2.66\times 10^{4}, 3.18\times 10^{4}, 3.8\times 10^{4}\right\}$ and $\left(x,y,z\right)= \left(0.1,0.012,1\right)$. Solid lines and markers represent the propagative low-$\Rm$ model and measurements respectively. White markers ($\vartriangleright,\square,\vartriangleleft$) highlight the cases at $\Ja = 0.1, 0.5$ and 1 respectively, all at $\Ha = 3.8\times 10^4$, studied in more detail on figures \ref{fig:mapping_orientation_angle_Ha_3p8e4_diff_Luw}, \ref{fig:mapping_alpha_exp_th} and \ref{fig:mapping_phase_shift_exp_th}. {In (b), the oscillating diffusive regime is represented by the white area and the propagative regime by the grey area. The hatched area depicts the range of $\Ja$ values where nonlinearities are observed.}}
\end{figure} 
\begin{figure}[h!]
	\begin{subfigure}[c]{0.5\linewidth}
		\centering
		\includegraphics[width=.97\linewidth]{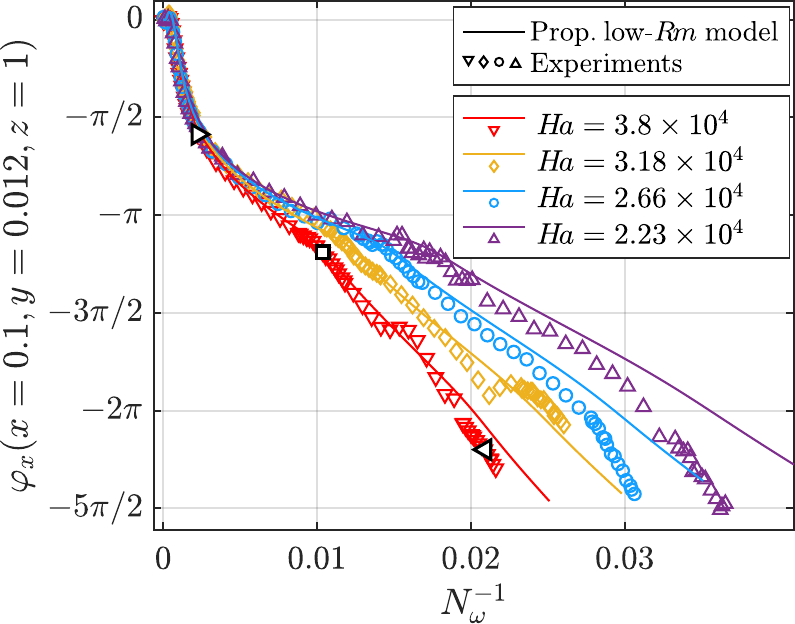}
	\caption{}\label{fig:phase_shift_vs_Nomega_exp_th}
\end{subfigure}
\hfill
\begin{subfigure}[c]{0.5\linewidth}
	\centering
	\includegraphics[width=.97\linewidth]{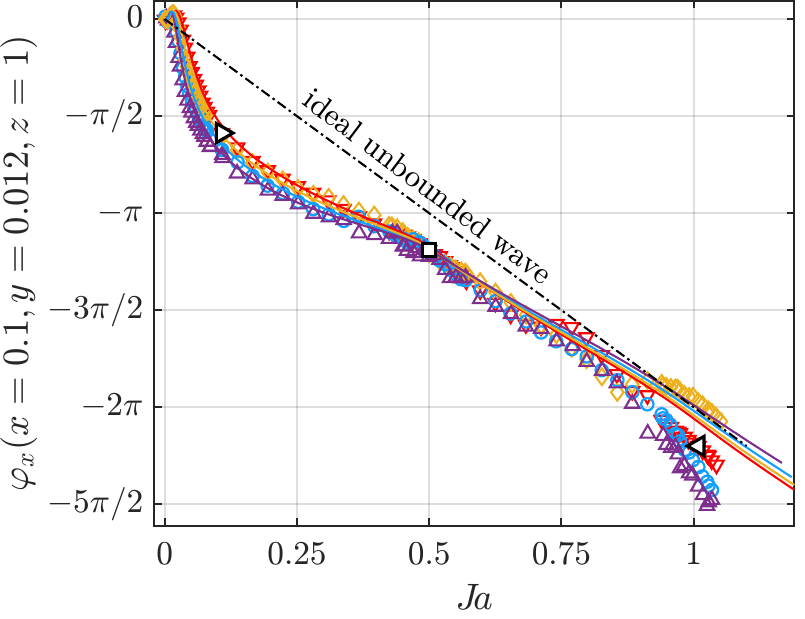}
	\caption{\label{fig:phase_shift_vs_LuOmega_exp_th}}
\end{subfigure}
\caption{\label{fig:phase_shift_vs_Lu_and_N_omega_exp_th} Phase shift $\varphi_{ x}$ against $\NOmega^{-1}$ (a) and $\Ja$ (b) for $\Ha = \left\{1.9\times 10^{4}, 2.66\times 10^{4}, 3.18\times 10^{4}, 3.8\times 10^{4}\right\}$ and at $\left(x,y,z\right)= \left(0.1,0.012,1\right)$. Solid lines and markers represent the propagative low-$\Rm$ model and measurements respectively. The dash\textendash dotted line (\protect\dashDottedLine) $\varphi_x$ is for an ideal unbounded wave.
White markers ($\vartriangleright,\square,\vartriangleleft$) highlight the cases at $\Ja = 0.1, 0.5$   and 1, respectively, all at $\Ha = 3.8\times 10^4$, studied in more detail on figures \ref{fig:mapping_orientation_angle_Ha_3p8e4_diff_Luw}, \ref{fig:mapping_alpha_exp_th} and \ref{fig:mapping_phase_shift_exp_th}. 
}
\end{figure} 
To track the existence of the diffusive and propagative flow regimes identified in $\S$ \ref{sb_sec:AW_characteristics}, we first analyse how these quantities vary with the screen parameter $\Rv$: figures \ref{fig:att_coef_vs_Lu_and_N_omega_exp_th} and \ref{fig:phase_shift_vs_Lu_and_N_omega_exp_th} show $\alpha_x(x=0.1,y=0.012,z=1)$ and $\varphi_{ x}(x=0.1,y=0.012,z=1)$ from both theory and experiments, plotted against $\Ja$ and $\NOmega^{-1}$.
The variations of both $\alpha_x$ and $\varphi_{ x}$ are very similar to those observed for the single-electrode model in $\S$ \ref{sec:ac_waves}. The attenuation coefficient exhibits a plateau in the limit $N_\omega^{-1}\rightarrow0$, a maximum around $\NOmega^{-1} = 1 \times 10^{-3}$, followed by a sharp decay at higher values, with a second peak in the higher frequencies. 
Similarly, the phase shift $\varphi_{ x}$ displays two distinct sequences of monotonic decay at low and high screen parameter $\Rv$, which are qualitatively very similar to those observed for a single electrode.
The collapse of both experimental and theoretical data into a single curve for $\NOmega^{-1} \lesssim 5\times10^{-3}$ in figures \ref{fig:att_coef_vs_Lu_and_N_omega_exp_th}$(a)$ and \ref{fig:phase_shift_vs_Lu_and_N_omega_exp_th}$(a)$ confirms that the variations of $\alpha_x$ within this range reflect the spatial attenuation of diffusive oscillations. Similarly, the alignment of resonant peaks at $\Ja=0.5$ and 1 on figure \ref{fig:att_coef_vs_Lu_and_N_omega_exp_th}$(b)$ is a clear signature of resonant waves characteristic of the propagative regime. {This point is one of the main results of this work as it identifies both experimentally and theoretically the two parameters $\NOmega$ and $\Ja$ that, respectively, control the purely diffusive regime and the regime where MHD waves akin to AW can be electrically driven at low-magnetic Reynolds number, as well as the transition between these regimes.}

The agreement between theory and experiment is generally excellent, but two 
distinct types of discrepancies deserve further comments.
First, an offset separates the collapsed theoretical from the collapsed experimental values of $\alpha_x$ (figure \ref{fig:att_coef_vs_Lu_and_N_omega_exp_th}) resulting in a $\simeq\!15\%$ difference in the value of the peak in the diffusive regime and $\simeq 15\%$ in its position.  For the propagative peaks, these discrepancies are, respectively, of $\simeq 10\%$ and $\simeq 3\%$. However, the experimental data still shows excellent collapse in the diffusive regime and the propagative peak are all well aligned to $\Ja=0.5$ and 1. Further calculations not reported here show that the actual value of $\alpha_x$ is quite sensitive to the choice of distribution of injected current across the electrode $j^w$. While our simplified model of a Gaussian distribution  (\ref{eq:jw_gauss}) is numerically convenient and roughly reflects the localisation of the injected current at an electrode, it is very different from the complex distribution of current that occurs in the experiment as a result of the interaction between the liquid phase and the solid conductor that forms the electrode \citep{herreman_2019_prf}. Hence, this offset can be attributed to our simplified model of the current distribution and we argue that this minor difference does not detract from the excellent agreement between the physical mechanisms taking place in FlowCube and those captured by the model.\\
The second discrepancy takes place in the high $\Rv$ limit, within the propagative regime: for $\Ja >0.85$, both the attenuation and phase shift depart significantly in value and in behaviour from the model. The excellent agreement between theory and model elsewhere suggests that the model misses a physical mechanism in this regime. Since waves propagate more freely in this regime, they reach higher amplitudes and may thus interact nonlinearly, a process that the linear model cannot capture and that {is discussed further in the next sections.}  
\begin{figure}[h!]
\begin{tabular}{|c|}
	\hline	
	\begin{subfigure}[c]{1\linewidth}
	\centering
	(a)\\
	\includegraphics[width=1\linewidth]{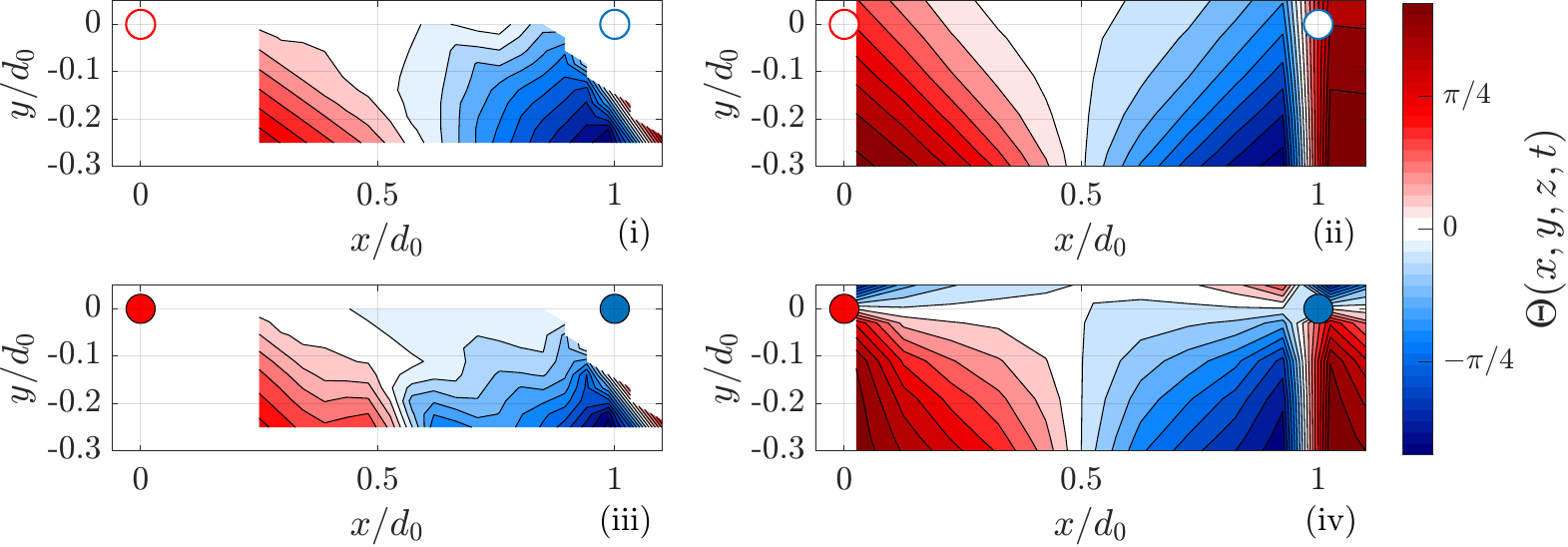}
	\end{subfigure}\\
	\hline
	\begin{subfigure}[c]{1\linewidth}
	\centering
	(b)\\
	\includegraphics[width=1\linewidth]{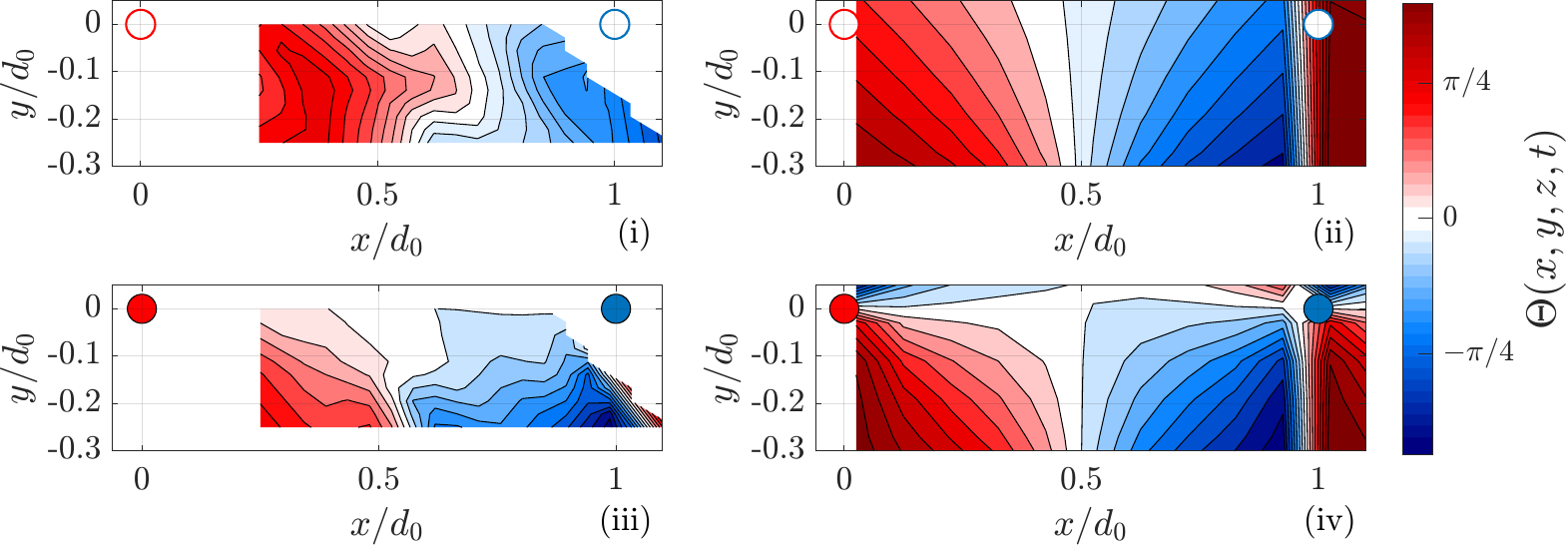}
	\end{subfigure}\\
	\hline
	\begin{subfigure}[c]{1\linewidth}
	\centering
	(c)\\
	\includegraphics[width=1\linewidth]{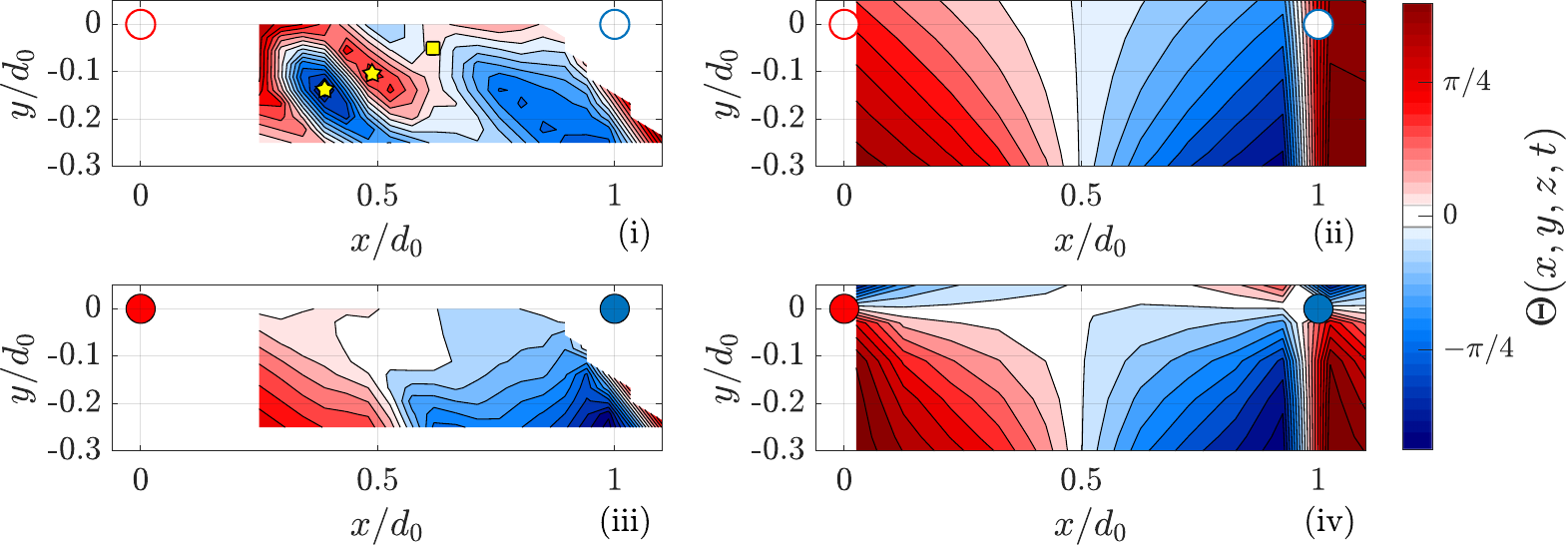}
	\end{subfigure}\\
	\hline
\end{tabular}
\caption{\label{fig:mapping_orientation_angle_Ha_3p8e4_diff_Luw} Snapshot contours of $\Theta$ for $\Ja = 0.1$ $(a)$, $\Ja = 0.5$ $(b)$ and $\Ja = 1$ $(c)$. Here $\Ha= 3.8\times 10^4$ for all cases. For each value of $\Ja$, instantaneous contours of $\Theta$ are plotted at the top wall (i,ii) and at the bottom wall (iii,iv), from experimental data (i,iii) and from the {propagative low-$\Rm$} model (ii,iv). The full coloured circles show the location of two of the four injection electrodes --  in phase opposition -- while the empty circles show their virtual projection on the top wall. The yellow stars {and the square} highlight the focus points {and the saddle respectively}. {The time considered for snapshots is arbitrary. However, other times give similar results.}}
\end{figure} 
\subsection{Linear and nonlinear flow patterns}
Having identified the diffusive and propagative regimes in the experimental data, {we now seek} to understand the discrepancy between theory and experiments at high values of the Jameson number. To this end, the topology of the flow is investigated using {the angle between the potential gradients and $\unitx$} at the Hartmann walls, $\Theta(z=0)$ and $\Theta(z=1)$. This quantity has the advantage of ignoring the amplitude of the potential gradient and so reflects solely the topology of the flow. We focus on three different cases: one in the oscillating diffusive regime, one well captured by the model within the propagative regime and one within the propagative regime but with significant discrepancy between model and experiment. We set $\Ha= 3.8\times 10^{4}$ for all three cases corresponding to $\Ja= \left\{0.1, 0.5, 1\right\}$, respectively. These cases are highlighted by three different white-filled symbols in figures \ref{fig:att_coef_vs_Lu_and_N_omega_exp_th} and \ref{fig:phase_shift_vs_Lu_and_N_omega_exp_th}.
Instantaneous experimental and theoretical contours of $\Theta$ at the top and bottom walls ($z=0$ and $z=1$) are plotted in figure \ref{fig:mapping_orientation_angle_Ha_3p8e4_diff_Luw} for all three cases. 

For $\Ja=0.1$ and 0.5 (figures \ref{fig:mapping_orientation_angle_Ha_3p8e4_diff_Luw}$(a)$ and \ref{fig:mapping_orientation_angle_Ha_3p8e4_diff_Luw}$(b)$), isolines of $\Theta$ from experiments are topologically consistent with those from the model. The main difference 
is that positive isovalues of $\Theta$ swell slightly towards the positive $x$-direction in the experimental data at $\Ja= 0.5$. 
While this could be due to nonlinearities, the limited spatial resolution in the experiments and the added uncertainty of comparing instantaneous snapshots make it difficult to reach a definite conclusion regarding the origin of this effect based on these snapshots alone.

For $\Ja=1$, the experimental contours of $\Theta$ are consistent with theory at the bottom wall $z=0$. At the top wall $z=1$, by contrast, two focus points and at least one saddle appear at $\left(x,y\right)= \left(0.38,-0.14\right)$, $\left(x,y\right)= \left(0.48,-0.11\right)$ {and $\left(x,y\right)= \left(0.62,-0.03\right)$}, respectively (figure \ref{fig:mapping_orientation_angle_Ha_3p8e4_diff_Luw}c). Further critical points must exist on the top wall to satisfy the topological constraints on their numbers \citep{hunt1978_jfm} but these fall outside of the spatial visualisation window available to us. Because of these, the topology found in the experimental data differs fundamentally from that predicted by the model. This new topological structure is not compatible with the linear prediction, and implies an underpinning nonlinear mechanism. 
Furthermore, the two focus points are located in the region of the isoline swelling observed for the case $\Ja=0.5$, suggesting that the topological change may ensue from small nonlinear deviations to the linear prediction at lower values of $\Ja$. This also suggests that nonlinearities are favoured at higher values of $\Ja$ and especially near the resonance $\Ja=1$, {i.e.} further into the propagative regime, and where the amplitude of the waves is greatest. These conditions are most favourable to the occurrence of nonlinearities and the resonance points to a $\Ja$ nonlinear self-interaction of the wave.
{Generally speaking, nonlinearities may potentially arise out of nonlinear wave interaction but also from hydrodynamic instabilities of the oscillating flow not involving wave interaction}: instabilities of forced electromechanical oscillations were indeed observed, but at $\Ha<10$, i.e. outside of the propagative regime of relevance here \citep{thomas2010_jfm}. 
{Here, however, the values of the measured Reynolds number at the bottom wall $\Rey^{\rm bot} = \{199, 102, 75\}$ and at the top wall $\Rey^{\rm top} = \{ 48, 8, 8\}$ for $\Ja = \{0.1,0.5,1\}$, respectively, are very low. Furthermore, the strongest changes on the flow topology are observed at the top wall for $\Ja =1$, which is where the value of $\Rey$ is the lowest. Hence,  hydrodynamic instabilities are unlikely to be the cause of the instabilities.}
 
In any case, this change in topology corresponds to the discrepancy observed on \S \ref{sb_sec:two_regimes_exp_and_th} for $\Ja>0.85$ and so confirms that these discrepancies reflect a nonlinear regime of AW, not accounted for in the model. 
\subsection{Influence of horizontal gradients on the diffusion of oscillations and wave propagation}\label{sb_sec:horizontal_gradient_exp}
\begin{figure}[h!]
	\centering
	\includegraphics[width=1\linewidth]{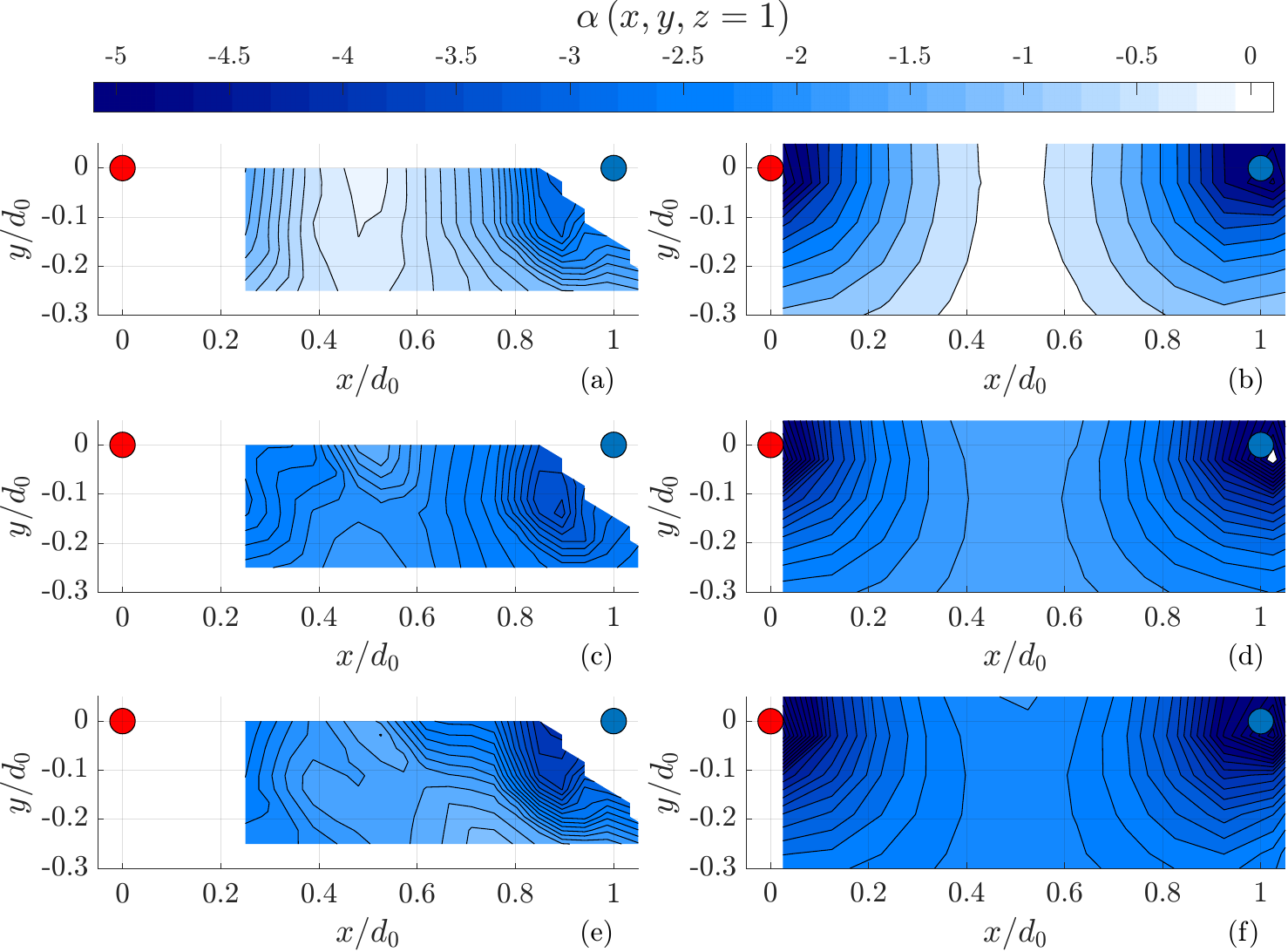}
\caption{\label{fig:mapping_alpha_exp_th} Snapshot contours of $\alpha$ at the top wall ($z=1$) for $\Ja = 0.1$ $(a,b)$, $\Ja = 0.5$ $(c,d)$ and $\Ja = 1$ $(e,f)$. Here $\Ha= 3.8\times 10^4$ for all cases. For each value of $\Ja$, contours of $\Theta$ are plotted from experimental data $(a,c,e)$ and from the {propagative low-$\Rm$} model $(b,d,f)$. The coloured circles show the location of two electrodes in phase opposition out of the four injection electrodes.}
\end{figure} 

\begin{figure}[h!]
	\centering
	\includegraphics[width=1\linewidth]{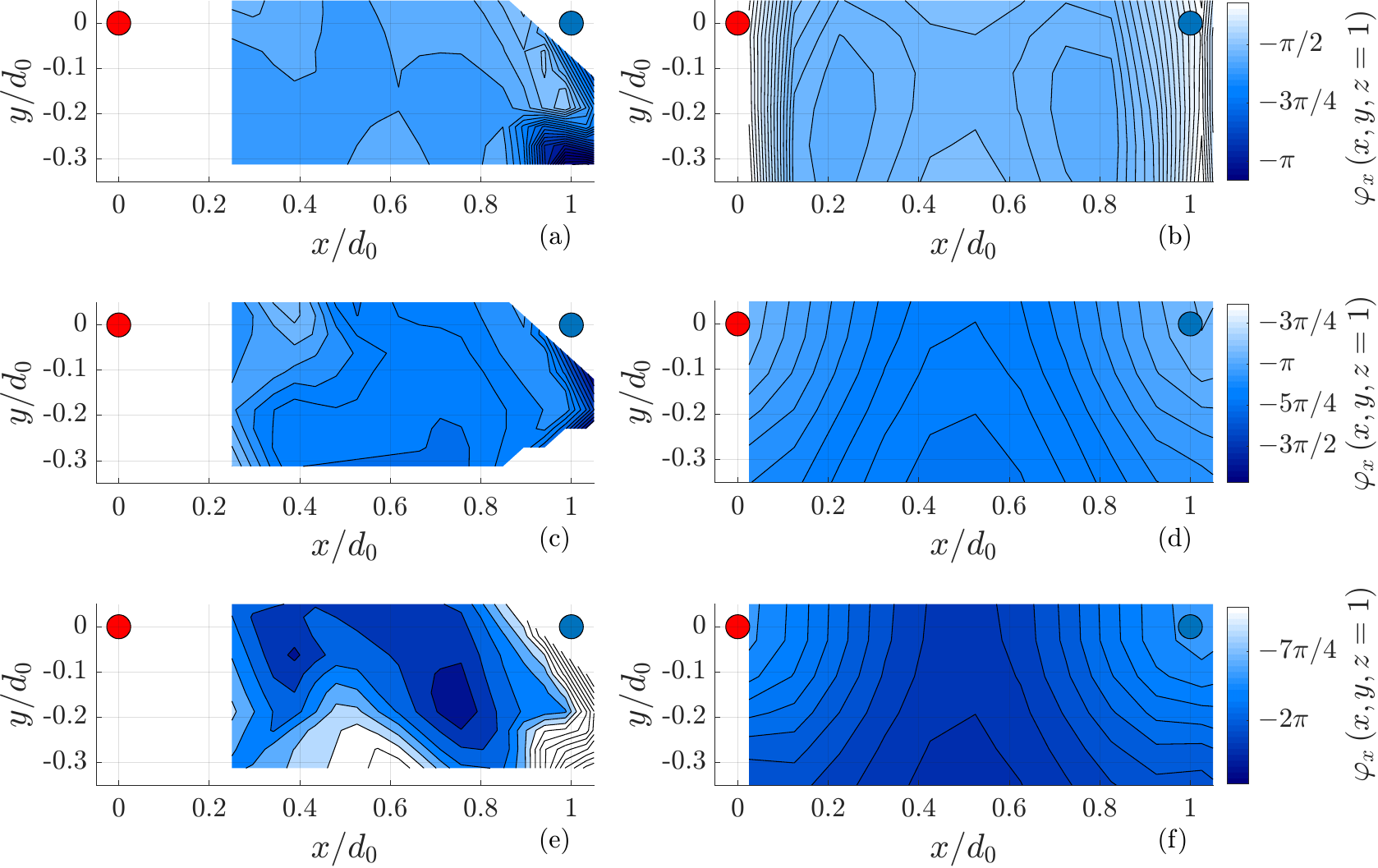}
\caption{\label{fig:mapping_phase_shift_exp_th} Snapshot contours of $\varphi_{ x}$ at the top wall ($z=1$) for $\Ja = 0.1$ $(a,b)$, $\Ja = 0.5$ $(c,d)$ and $\Ja = 1$ $(e,f)$. Here $\Ha= 3.8\times 10^4$ for all cases. For each value of $\Ja$, the contours of $\Theta$  are plotted from experimental data $(a,c,e)$ and from the {propagative low-$\Rm$} model $(b,d,f)$. The coloured circles show the location of two electrodes in phase opposition out of the four injection electrodes.}
\end{figure} 
A key result of the theoretical study in \S \ref{sb_sec:AW_characteristics} was that both the diffusion of oscillations and wave propagation are influenced by the transverse gradients of electric potential. We shall now track and quantify this effect experimentally. The measurements in FlowCube deliver attenuation and phase shifts between the Hartmann walls ($z=0$ and $z=1$) only. Hence, we plot the experimental and theoretical contours of these quantities, $\alpha(x,y,z=1)$  and $\varphi_{ x}(x,y,z=1)$, on figures \ref{fig:mapping_alpha_exp_th} and \ref{fig:mapping_phase_shift_exp_th}, respectively, again focusing on {the value of $\Ha$} displaying the most prominent wave propagation at high $\Ja$, i.e. $\Ha = 3.8 \times 10^4$ and for $\Ja \in\{0.1, 0.5, 1\}$.

Experimental contours of $\alpha$ and $\varphi_{ x}$ exhibit topologically equivalent patterns to their theoretical counterparts  except for $\Ja= 1$ (within the same limitations on the spatial resolution in the experiment and the error incurred by comparing snapshots as in the previous section). The gradients of these quantities visibly decrease with the distance to the neighbouring electrode in all cases but the experimental one at $\Ja = 1$ too. The value of $\alpha$ itself (figure \ref{fig:mapping_alpha_exp_th}) increases with the transverse distance form each electrode, both in the oscillating diffusive regime (figure \ref{fig:mapping_alpha_exp_th}a-b) and in the linear propagative regime (figure \ref{fig:mapping_alpha_exp_th}c-f) in both theory and experiment. 
Experimental contours of $\varphi_{ x}$ are consistent with the model for $\Ja=0.1$ and 0.5 (figure \ref{fig:mapping_phase_shift_exp_th}). In both cases, the magnitude of $\varphi_{ x}$ increases with the distance to the neighbouring electrode. These observations concur to show that the high transverse gradients in the vicinity of the electrodes increase the attenuation of the oscillations but accelerate their diffusion and their propagation along $\unitz$. 

For $\Ja=1$, the experimental contours of both $\alpha$ and $\varphi_{ x}$ deviate from the model. This case, highlighted by marker ($\vartriangleleft$) on figures \ref{fig:att_coef_vs_Lu_and_N_omega_exp_th} and \ref{fig:phase_shift_vs_Lu_and_N_omega_exp_th}, corresponds to the regime where nonlinear wave interactions incur a change in the flow topology at the top wall (figure \ref{fig:mapping_orientation_angle_Ha_3p8e4_diff_Luw}c), a phenomenon that the linear theory cannot capture. In the experimental data, the locus of the emerging nonlinear pattern coincides with a locally enhanced phase shift compared with the model, indicating that the nonlinear pattern travels along $\unitz$ at a different velocity from the fundamental mode driven by the forcing.\\ 

To further quantify the dependence of the attenuation and phase shift on the horizontal gradients of electric potential, we take advantage of the near-monotonic dependence of these gradients on the transverse distance to the nearest electrode  in the $\left(x,y\right)$ plane, which we denote $r^*$. Based on this, the lengthscale of transverse gradients can be approximated by $r^*$. We focus on the linear propagative regime where experiments and theory agree well, {e.g.} $\Ja=0.5$. Figures \ref{fig:alpha_vs_rn} and \ref{fig:phase_shift_gradx_phi_vs_rn}, respectively, show the variations of $\alpha$ and $|\varphi_{ x}|$ with $r^*$ from the experiment and the model. Since these fields are not axisymmetric about either of the electrodes, these variations also depend on the azimuthal angle with respect to the closest electrode, especially for points that are nearly equidistant from two electrodes. The theoretical range of variation of $\alpha$  and $\varphi_{ x}$ spanned in this way is represented on the figures by a red zone. The experimental data shows a greater spread attributed to the local measurement error, of around $4\%$, and to the offset in damping rate incurred by our simplified model of the current distribution at the electrode (see $\S$ \ref{sb_sec:two_regimes_exp_and_th}).
Nevertheless, the experimental values of $\alpha$ are consistent with the model. In both cases, $\alpha$ increases with $r^*$. These variations do not appear to follow a simple scaling law. Indeed, the values of $r^*$ fall in an intermediate range between the regime of low transverse gradient investigated by \cite{jameson_1964} and a regime of much higher transverse gradient where waves are highly damped. In the high damping regime, horizontal gradients are sufficiently strong to incur magnetic dissipation comparable to the axial ones, and so considerably increase the damping of oscillations along $\unitz$. This phenomenon takes place in the deep blue area very close to the electrode in figure \ref{fig:mapping_att_coef_Ha_Reta}(c). The vertical contours of $\alpha$ obtained from the single electrode theory show that this region extends over the entire height of the vessel. In the QSMHD limit, this region becomes the classical viscous vortex core of inertialess thickness $\sim \Ha^{1/2}$ studied in detail by \cite{sommeria1988_jfm}.
\begin{figure}[h]
	\centering
	\includegraphics[width=0.6\linewidth]{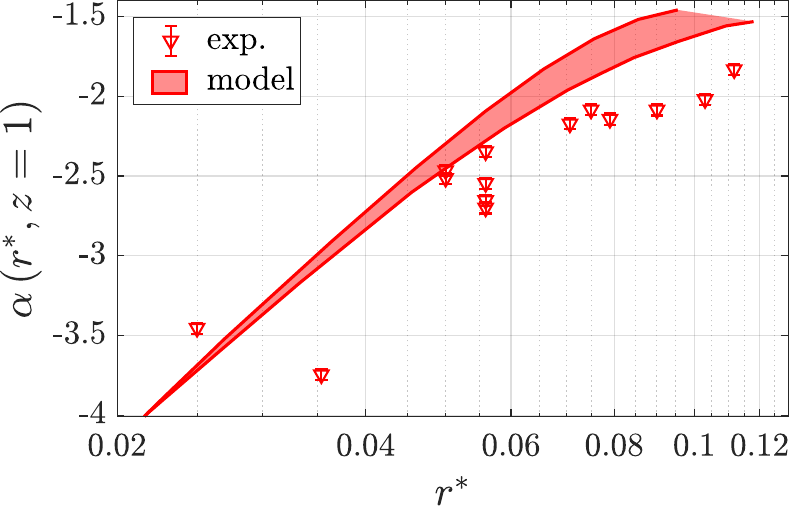}
	\caption{\label{fig:alpha_vs_rn} Here $\alpha$ versus $r^*$ for $\Ja =0.5$ and $\Ha= 3.8\times 10^{4}$. Here $r^*$ is the distance in the $(x,y)$ plane between a given measurement location and the nearest injection electrode. The solid line and markers correspond to the propagative low$-\Rm$ model and measurements, respectively. The error bars show the local measurements errors for $\alpha$, of $4\%$.}
\end{figure} 

Similarly, the magnitude of the phase shift $\varphi_{ x}$ increases with $r^*$ in both experiment and model, without following any obvious scaling law (figure \ref{fig:phase_shift_gradx_phi_vs_rn}). It should {also be noticed} that {the experimental data is more scattered} for $|\varphi_{ x}|$ than for $\alpha$. The scattering can be attributed to several factors: first, using a single component of the potential gradient $\varphi_x$ incurs errors when the gradient is close to normal to $\unitx$; second, the most scattered points fall mostly in the region where discrepancies in flow topologies were observed between theory and experiment. This suggests that the scattering may arise out of nonlinear effects.

The variations of $\alpha$ and $\varphi$ quantify the previous observation that stronger transverse gradients damp the waves whilst accelerating their propagation. Additionally, they show that the dissipative waves observed in FlowCube sit at an intermediate regime of dissipation. As such, less dissipative waves may be obtained in a different geometry minimising the transverse gradient. \cite{jameson_1964} {carefully positioned his measurement probe specifically to target this range. In doing so, he maximised the intensity of the waves he was able to} observe within the physical limitations of his experiment and despite the relatively low magnetic field available to him. The quality of the results he obtained in this manner, compared with other experiments available in these times \citep{lundquist_1949, bostick1952_pre,lehnert_1954,allen1959_prl} is a tribute to his deep understanding of the subtleties of these waves.
\begin{figure}[h]
\centering
\includegraphics[width=0.6\linewidth]{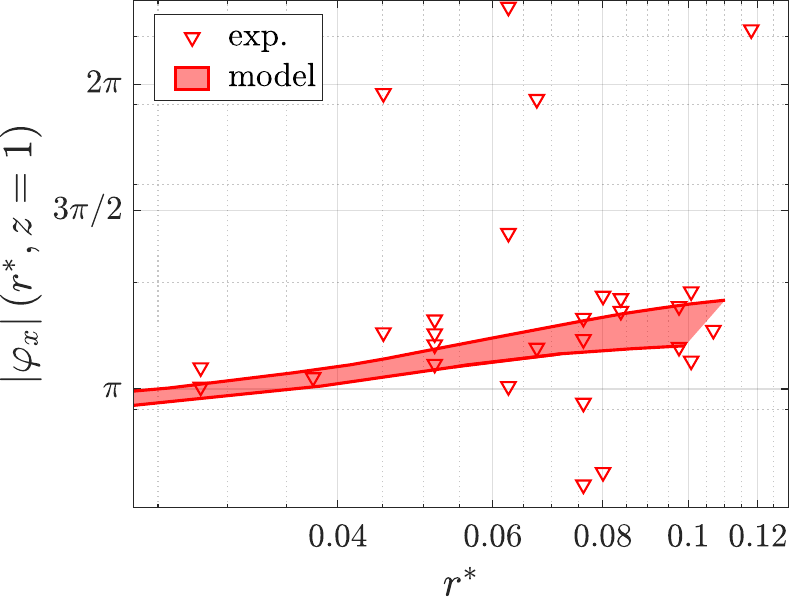}
\caption{\label{fig:phase_shift_gradx_phi_vs_rn} Here $\varphi_{ x}$ against $r^*$ for $\Ja =0.5$ and $\Ha= 3.8\times 10^{4}$. Here $r^*$ is the distance in the $(x,y)$ plane between a given measurement location and the nearest injection electrode. The solid line and the light red coloured area represent {the solution of the propagative low-$\Rm$} model while markers show measurements.}
\end{figure} 

\subsection{Linearity and nonlinearity of the experimental waves}
The discrepancies between model and experiment at high Jameson numbers ($ \Ja \gtrsim0.85$) in \S\ref{sb_sec:two_regimes_exp_and_th} revealed significant nonlinearities in this regime. Smaller discrepancies between theoretical and experimental contours of the potential gradient angle further suggest that these 
nonlinearities develop at lower Jameson numbers ($\Ja\simeq0.5$). Hence, notwithstanding the otherwise excellent agreement between theory and experiment, {the question arises {of which range of control parameters sees these} nonlinearities {affect} either the oscillating diffusive regime or the propagative one}.
In particular, how much of the experimental data may fall within this regime? To answer these questions, we assess the linearity of electric potential gradients with respect to the forcing intensity. Dimensionally, the forcing intensity is controlled by the total current injected in the flow, and non-dimensionally, it is measured by $\Rey_0$. 
Nonlinearities are assessed using the average intensity of the electric potential gradients, and their relative nonlinearity is defined as
\begin{equation}
	\epsilon_{\rm NL}(z)=\left| 1-\frac{\left\langle \left|\widehat{\bm\nabla_x\phi}(\Rey_0)\right|\right\rangle_{x,y}}{\Rey_0} \lim_{\Rey_0\rightarrow0}{\frac{\Rey_0} {\left\langle \left|\widehat{\bm\nabla_x\phi}(\Rey_0)\right|\right\rangle_{x,y}}}\right|,
\end{equation}
where $|\widehat{\bm\nabla_x\phi}|$ refers to the amplitude of the oscillations of $\partial_x\phi$.  
{The quantity $\epsilon_{\rm NL}(z)$ can be understood as the relative discrepancy averaged at either wall ($z=0$, or $z=1$) between the measured electric potential gradients at a finite $\Rey_0$ and their value expected from linear upscaling of measurements made in the linear limit $\Rey_0\rightarrow0$. Hence,}
for flows purely within the linear regime, $\epsilon_{\rm NL}=0$. To assess this quantity, the purely linear solution in the limit $\Rey_0\rightarrow0$ is approximated by the experimental data obtained at the lowest value of $\Rey_0$, \ie $\Rey_0= 3.75 \times 10^3$ (for which $\Rey^{\rm bot}= 30.0$  and $\Rey^{\rm top}= 7.7$ at $\Ja = 0.1$ and $\Ha = 3.8\times 10^4$). 
The relative nonlinearity of the potential gradient $\epsilon_{\rm NL}$ is plotted at the bottom and top walls for $\Ha =\left\{3.18\times 10^4, 3.8\times 10^4\right\}$ and $\Ja= \left\{0.1, 0.5, 1\right\}$ (figure \ref{fig:top_bot_ampl_mean_vs_Re0}). 
\begin{figure}[h]
	\begin{subfigure}[c]{1\linewidth}
		\centering
		\includegraphics[width=0.65\linewidth]{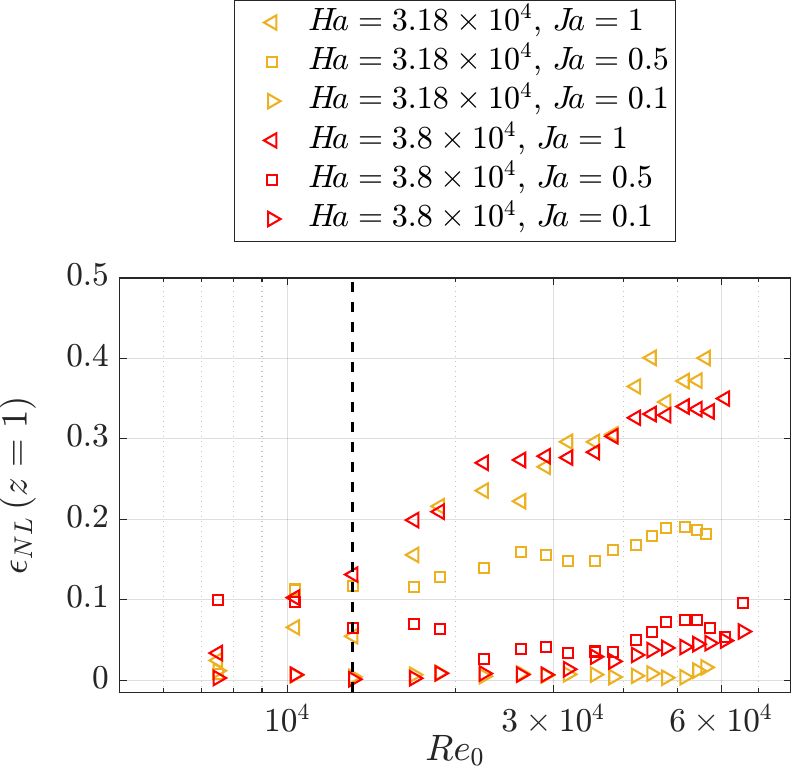}
	\caption{}\label{fig:top_ampl_mean_vs_Re0}
	\vspace{2mm}
\end{subfigure}
\begin{subfigure}[c]{1\linewidth}
	\centering
	\includegraphics[width=.65\linewidth]{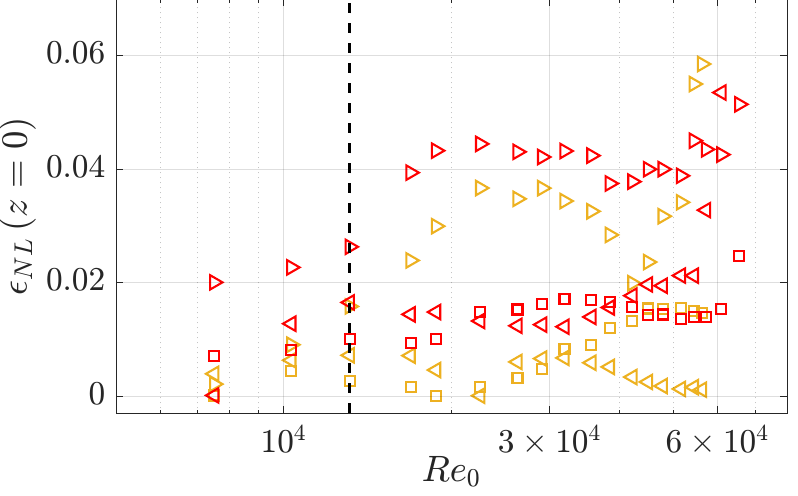}
	\caption{\label{fig:bot_ampl_mean_vs_Re0}}
\end{subfigure}
	\caption{\label{fig:top_bot_ampl_mean_vs_Re0} Normalised $\left\langle |\widehat{\bm\nabla_x\phi}|\right\rangle$ against $\Rey_0$ at the top wall (a) and the bottom wall (b) for $\Ha =\left\{3.18\times 10^4, 3.8\times 10^4\right\}$ and $\Ja= \left\{0.1, 0.5, 1\right\}$. For each case of the set $\left\{\Ha, \Ja\right\}$, $\left\langle |\widehat{\bm\nabla_x\phi}|\right\rangle$ is normalised by its value at the lowest experimental $\Rey_0$.}
\end{figure} 
In all cases, $\epsilon_{\rm NL}$ smoothly increases with $\Rey_0$. Thus, nonlinearities are always present, albeit in a vanishingly small amount in the limit $\Rey_0\rightarrow0$. This concurs with the observation of small discrepancies in the contours of the angle of electric potential gradients $\Theta$ (figure \ref{fig:mapping_orientation_angle_Ha_3p8e4_diff_Luw}b-c), even in regimes where theoretical values of both the attenuation coefficient and phase shift agree well with the experimental ones.\\
At the bottom wall, $\epsilon_{\rm NL}$ remains below $6\%$ for all cases (figure \ref{fig:bot_ampl_mean_vs_Re0}). A saturation in the variations of $\epsilon_{\rm NL}(\Rey_0)$, however, takes {place} around $\Rey_0\simeq2\times10^4$, and only in the propagative regime $\Ja\geq0.5$ (for $\left\{\Ha= 3.8\times 10^4, \Ja= 0.5\right\}$ and $\left\{\Ha= 3.8\times 10^4, \Ja= 1\right\}$; figure \ref{fig:bot_ampl_mean_vs_Re0}). This is indicative of a change in nonlinear dynamics with a possible dominance of nonlinear wave interaction. By contrast, flows within the diffusive regime remain essentially linear for the entire range of forcing spanned within the experiment. 
Data at the top wall (figure \ref{fig:top_ampl_mean_vs_Re0}) reveal the same phenomenology but with significantly greater nonlinearity, up to $\simeq40\%$. 
This also concurs with the observations of the contours of $\Theta$, where the discrepancy between theory and experiment was found greater {at the smallest $\Rey$ and at the largest $\Ja$ values}, near the top wall. 
This further supports the idea that nonlinearities arise from nonlinear wave interactions  {rather than from other hydrodynamics instabilities}. 
Indeed, these interactions are impeded near the bottom wall where patterns of injected current are imposed. However, they grow as they propagate away from the forcing region, and so are more likely observed near the top wall.

Except for the present section, all experimental data shown in this work was obtained for $\Rey_0=1.33\times10^4$ (highlighted with  dashed lines on figures \ref{fig:top_ampl_mean_vs_Re0} and \ref{fig:bot_ampl_mean_vs_Re0}). For this forcing intensity, $\epsilon_{\rm NL}$ remains lower than $\simeq12\%$ for all cases but $\Ha= 3.8\times 10^4, \Ja= 1$. Hence, it can be concluded that the phenomenology discussed throughout this work is linear in all regimes where experiment and theory agree. Nevertheless, the data at higher $\Rey_0$ discussed in this section clearly shows that nonlinear regimes are easily reached within FlowCube's operational range of parameters. 

\section{Conclusion}
We studied an electrically driven oscillating flow confined between two walls perpendicular to an externally imposed magnetic field, {in order to identify} the conditions in which diffusive, propagative and nonlinear processes occur in low-$\Rm$ MHD. The underlying question at stake is whether  AW observed in astrophysical and geophysical systems can be studied in detail in laboratory-scale experiments using liquid metals. To address this problem, we developed two linear models, one based on a propagative extension of the usual low-$\Rm$ MHD approximation, and one based on the QSMHD approximation. We also adapted the FlowCube experimental device \citep{klein_potherat_2010} to directly track these regimes in a rectangular vessel filled with liquid metal. The propagative low-$\Rm$ approximation differs from the usual QSMHD approximation in that the time scale of local oscillations {of momentum and induced magnetic field} is different from the flow turnover time. It is therefore governed by three non-dimensional parameters: the usual Hartmann number $\Ha=Bh(\sigma/(\rho\nu)^{1/2})$ and two screen parameters, a viscous one, $\Rv= \omega h^2/(2\pi \nu)$ and a resistive one $\Rn=\omega h^2/(2\pi \eta)$. The linearised QSMHD approximation is recovered in the limit $\Rn\rightarrow0$, keeping $\Rv$ finite. {This approach enabled us to keep the induction time scale as a free parameter and so cover the full range of regimes where diffusion and propagation compete. This range can be mapped to other problems where the induction time scale is controlled by other processes such as advection or convection, and so offers a general framework for Alfv\'en waves at low magnetic Reynolds number.}

Combining theory and experiments enabled us to identify and characterise three different regimes of forced oscillations.  
The first regime falls within the QSMHD approximation. It is captured by both  linear models and experiments and occurs at low screen parameters $\Rv$ or $\Rn$. In this regime, the Lorentz force acts exclusively so as to diffuse momentum along magnetic field lines. In the limit $\Rn\rightarrow0$, this diffusion process drives flows towards a quasi-two-dimensional state \citep{sommeria_why_1982}. Hence, we called this regime {oscillating diffusive}, and found that it is characterised by an oscillating parameter $\NOmega=2\pi \sigma B_0^2/{\rho\omega}$ built out of the ratio of the {oscillating acceleration} time scale $2\pi/\omega$  to the two-dimensionalisation time scale at the scale of the box associated with the Lorentz force, $\tau_{2D}=\rho/(\sigma B_0^2)$, first introduced by \cite{sommeria_why_1982}. 
The second regime occurs at higher values of $\Rn$. It is captured by experiments in excellent agreement with the linear propagative low-$\Rm$ model, but not by the QSMHD model. This linear {propagative} regime is dominated by the propagation of AW. Their resonance across the channel occurs at values of $\Rn$ such that the ratio
between the propagation time of Alfvén waves across the channel and the oscillation period is either 1/4, 1/2, 3/4 or 1. The second regime is characterised by values of this ratio, which we named the {Jameson number} $\Ja= \omega h / \left(2\pi V_A\right)$ below $\simeq0.85$.
The existence of this regime is one of the chief results of this study as it identifies the {regime in which MHD waves akin to AW, but different from ideal AW first theorised by Alfv\'en \citep{alfven_1942},} can propagate at low $\Rm$. 
The third regime occurs at the highest values of $\Rv$, corresponding to $\Ja\gtrsim0.85$ and therefore remains propagative in nature. It emerges where wave amplitudes are too high for the linear model to remain valid (most noticeably near the resonance $\Ja=1$). Indeed, experimental data departs from the linear model, and displays a clear nonlinear behaviour. Hence, we name this regime the {nonlinear propagative regime}.

The particular type of AW we found in the propagative regime, further exhibits unique propagation properties so far unobserved at low-$\Rm$. In the linear propagative regime, both the model and the experimental data display a dependence of the phase velocity on $\Rn$, so unlike ideal AW, these waves are dispersive. Furthermore, the configuration involves a spatially inhomogeneous forcing in the planes perpendicular to the magnetic field with electric potential gradients decaying away from the points of current injection. This provided us with an opportunity to analyse how such an inhomogeneity affects wave propagation. 
Indeed, both model and experiment clearly show that the propagation of these dispersive AW depends on the transverse gradients: AW are indeed both locally accelerated and more strongly damped in stronger gradients. {Such variations of propagation velocity, never observed before at low-$\Rm$, are sometimes referred to as {phase mixing}. It locally increases the viscous and resistive wave dissipation responsible for a heating process that may explain the anomalous temperature in the Sun's corona \citep{heyvaerts_1983,mclaughlin__2011_AA,prokopyszyn_2019_AA}}.

Outside the linear propagative regime, nonlinearities manifest themselves in several ways.
First, contour maps of electric potential reveal a change in the flow topology, noticeable near the top Hartmann wall, i.e. farther away from the forcing point. This, together with the occurrence near a resonance of these nonlinearities, supports the hypothesis of nonlinear self-interaction of AW. Additionally, varying the forcing intensity expressed by Reynolds number $\Rey_0$ for different values of $\Ja$ shows a stronger departure from the linear model at higher amplitudes, another signature of nonlinearity. The discrepancy from linearity was further found to increase continuously from very small $\Rey_0$ and so suggests that nonlinearities are always present, rather than ignited near a bifurcation as the forcing amplitude increases. The presence of nonlinearities driven by wave self-interaction is another major result and so answers one of the key questions that motivated this work: nonlinear energy transfers driven by waves can be reproduced in liquid metal experiments. This {opens new possibilities for the} study of a second mechanism directly relevant to processes expected to take place in astrophysical systems, especially the solar corona \citep{davila1987_apj,holst_2014}.

The three regimes found in this work show that three key ingredients of astrophysical and geophysical AW can be reproduced in a small-scale experiment: diffusion, propagation and nonlinearity. Additionally, we have been able to characterise the inhomogeneous properties of dispersive AW, another key feature of the complex dynamics of AW in these systems. These results open a new spectrum of opportunities to model such practically inaccessible objects as planetary interiors or the solar corona in the relative comfort of a small-scale laboratory, using liquid metals. The {AW in solar plasma or solar wind are unlikely to be fully reproduced in liquid metal experiments so we are certainly a very long way} from bringing the physics of the Sun into a box filled with liquid metal. But we are almost an equally long way from having exhausted the possibilities offered by the increasingly high intensity of magnetic fields available to liquid metal experiments, and the {flexibility} offered by electrically driven AW: varying the forcing intensity, the forcing geometry, the shape size of the box offer as many opportunities to seek more intense nonlinearities, different types of inhomogeneities and of nonlinear effects. Each of these may hold the key to some aspects of how AW behave in the Sun or in planetary interiors.

\noindent{\bf Acknowledgement} The authors are grateful to the European Magnetic Field Laboratory (EMFL) and the Laboratoire des Champs Magnétiques Intenses-Grenoble (LNCMI-Grenoble, part of theFrench CNRS) for providing us and for supporting access to their unique magnets with high magnetic field in sufficiently large bores to conduct fluid mechanics experiments.\\

\noindent{\bf Funding.} The authors would also like to thank the IDEX of Université Grenoble Alpes for its substantial contribution to the funding of Samy LALLOZ's doctoral scholarship through the International Strategic Partnerships (ISP) program. The SIMAP laboratory is part of the LabEx Tec 21 (Investissements d’Avenir, Grant Agreement No. ANR-11-LABX-0030). The UK Subscription to EMFL for access to the magnets at LNCMI-Grenoble is funded by EPSRC grant NS/A000060/1.\\

\noindent{\bf Author ORCIDs}\\
\orcidlink{0009-0001-5142-1302} Samy Lalloz \href{https://orcid.org/0009-0001-5142-1302}{https://orcid.org/0009-0001-5142-1302},\\
\orcidlink{0000-0002-7214-6821} François Debray \href{https://orcid.org/0000-0002-7214-6821}{https://orcid.org/0000-0002-7214-6821},\\
\orcidlink{0000-0001-9544-1578} Laurent Davoust \href{https://orcid.org/0000-0001-9544-1578}{https://orcid.org/0000-0001-9544-1578},\\
\orcidlink{0000-0001-8691-5241} Alban Pothérat \href{https://orcid.org/0000-0001-8691-5241}{https://orcid.org/0000-0001-8691-5241}.\\

\noindent {\bf Declaration of interests}. The authors report no conflict of interest.

\appendix
{
\section{Expression of coefficients $\left\{U_l^i,B_l^i\right\}_{l=1..8}$}\label{App:exp_for_U_B}
}
{
Here we detail the explicit expression of the unknown coefficients $\left\{U_l^i,B_l^i\right\}_{l=1..8}$, for the velocity and the magnetic disturbances, respectively. As a reminder, the solution $\kappa^i_{\perp}$ for the velocity and the magnetic disturbances is
\begin{align}
	\{u^i_{\theta}, \rmbt^i\}\left(z,t\right)&= \left\lbrace\mathrm{e}^{s_1^i\,z} \left(\left\lbrace U_1^i,B_1^i\right\rbrace \cos \left(t + \kappa_{1}^i\,z\right) - \left\lbrace U_2^i,B_2^i\right\rbrace \sin\left(t + \kappa_{1}^i\,z\right) \right)\right.\notag\\
	&+  \mathrm{e}^{-s_1^i\,z} \left(\left\lbrace U_3^i,B_3^i\right\rbrace \cos \left(t - \kappa_{1}^i\,z\right) - \left\lbrace U_4^i,B_4^i\right\rbrace \sin\left(t - \kappa_{1}^i\,z\right) \right) \notag\\
	&+ \mathrm{e}^{s_2^i\,z} \left(\left\lbrace U_5^i,B_5^i\right\rbrace \cos \left(t + \kappa_{2}^i\,z\right) - \left\lbrace U_6^i,B_6^i\right\rbrace \sin\left(t + \kappa_{2}^i\,z\right) \right) \notag \\
	&+ \left. \mathrm{e}^{-s_2^i\,z} \left(\left\lbrace U_7^i,B_7^i\right\rbrace \cos \left(t - \kappa_{2}^i\,z\right) - \left\lbrace U_8^i,B_8^i\right\rbrace \sin\left(t - \kappa_{2}^i\,z\right) \right) \right\rbrace,
	\label{App:eq:general_form_dist}
\end{align}
so that we define the coefficient vectors as $\bm B^i = \begin{bmatrix}  
	B_{l=1}^i &...& B_{l=8}^i
\end{bmatrix}^\mathrm{T}$ and $ \bm U^i = \begin{bmatrix}  
	U_{l=1}^i &...& U_{l=8}^i
\end{bmatrix}^\mathrm{T}$.
Here $\bm U^i$ and $\bm B^i$ are obtained using the boundary conditions (\ref{coef_BC_on_B_given_kt}, \ref{coef_BC_on_U_given_kt}), the Sturm\textendash Liouville problems (\ref{eq:Helmholtz_t_laplacian}, \ref{eq:Helmholtz_t_laplacian_z}) and the Navier-Stokes equation (\ref{eq:linear_ns_adm}). From these equations, we readily obtain the expression for $\bm U^i$,
\begin{equation}
	\bm U^i = {\begin{bmatrix} 
		\mathbfitsf{N}_1^i & \mathbfitsf{N}_2^i\\
		\Ha^{-2} {\mathbfitsf{\Dz}^i_1}^{-1} \left[\Rv \mathbfitsf\Dt-\mathbfitsf{M}_1^i \right] \mathbfitsf{N}_1^i &  \Ha^{-2} {\mathbfitsf\Dz^i_2}^{-1} \left[\Rv\mathbfitsf\Dt - \mathbfitsf{M}_2^i \right] \mathbfitsf{N}_2^i 
	\end{bmatrix}}^{-1}
	\begin{bmatrix} 
		\bm U_w^i\\
		\bm B_w^i\\	
	\end{bmatrix},\label{eq:BC_forcing}
\end{equation}
where are expressed the block-arrays for the magnetic and kinematic boundary conditions,
\begin{align}\label{eq:block_array_BC}
	\mathbfitsf{N}_1^i &= \begin{bmatrix}
		1 & 0 & 1 & 0\\
		0 & 1 & 0 & 1 \\
		\mathrm{e}^{s_1^i}\cos\kappa_{1}^i & -\mathrm{e}^{s_1^i}\sin\kappa_{1}^i &
		\mathrm{e}^{-s_1^i}\cos\kappa_{1}^i &	\mathrm{e}^{-s_1^i}\sin\kappa_{1}^i \\
		- \mathrm{e}^{s_1^i}\sin\kappa_{1}^i & -\mathrm{e}^{s_1^i}\cos\kappa_{1}^i&
		\mathrm{e}^{-s_1^i}\sin\kappa_{1}^i & -\mathrm{e}^{-s_1^i}\cos\kappa_{1}^i \\
	\end{bmatrix},\notag\\
	\mathbfitsf{N}_2^i &= \begin{bmatrix}
		1 & 0 & 1 & 0\\
		0 & 1 & 0 & 1 \\
		\mathrm{e}^{s_2^i}\cos\kappa_{2}^i & -\mathrm{e}^{s_2^i}\sin\kappa_{2}^i &
		\mathrm{e}^{-s_2^i}\cos\kappa_{2}^i &	\mathrm{e}^{-s_2^i}\sin\kappa_{2}^i \\
		- \mathrm{e}^{s_2^i}\sin\kappa_{2}^i & -\mathrm{e}^{s_2^i}\cos\kappa_{2}^i&
		\mathrm{e}^{-s_2^i}\sin\kappa_{2}^i & -\mathrm{e}^{-s_2^i}\cos\kappa_{2}^i \\
	\end{bmatrix},
\end{align}
the space first derivatives,
\begin{align}\label{eq:block_array_space_derivative}
		\mathbfitsf{\Dz}^i_1 &= \begin{bmatrix}
		s_1^i & -\kappa_{1}^i & 0 & 0\\
		\kappa_{1}^i & s_1^i & 0 & 0\\
		 0 & 0 & -s_1^i & \kappa_{1}^i\\
		 0 & 0 & -\kappa_{1}^i & -s_1^i\\
	\end{bmatrix},\;
		\mathbfitsf{\Dz}^i_2 = \begin{bmatrix}
		s_2^i & -\kappa_{2}^i & 0 & 0\\
		\kappa_{2}^i & s_1^i & 0 & 0\\
		0 & 0 & -s_2^i & \kappa_{2}^i\\
		0 & 0 & -\kappa_{2}^i & -s_2^i\\
	\end{bmatrix}, 
\end{align}
the time first derivative,
\begin{equation}\label{eq:block_array_time_derivative}
	\mathbfitsf\Dt = \begin{bmatrix}
		0 & -1 & 0 & 0\\
		1 & 0 & 0 & 0\\
		0 & 0 & 0 & -1\\
		0 & 0 & 1 & 0\\
	\end{bmatrix},
\end{equation}
 and the vector Laplacian operator,
\begin{align}\label{eq:block_array_Laplacian}
	\mathbfitsf{M}_1^i &= \begin{bmatrix}
		{s_1^i}^2 - {\kappa_{1}^i}^2 -{\kappa_{\perp}^i}^2& -2\kappa_{1}^i s_1^i& 0 & 0\\
		2\kappa_{1}^i s_1^i &  {s_1^i}^2 -{\kappa_{1}^i}^2 - {\kappa_{\perp}^i}^2 & 0 & 0\\
		0 & 0 &{s_1^i}^2 - {\kappa_{1}^i}^2 - {\kappa_{\perp}^i}^2& -2\kappa_{1}^i s_1^i\\
		0 & 0 & 2\kappa_{1}^i s_1^i &  {s_1^i}^2 - {\kappa_{1}^i}^2 - {\kappa_{\perp}^i}^2\\
	\end{bmatrix},\notag\\
	\mathbfitsf{M}_2^i &= \begin{bmatrix}
		{s_2^i}^2 -{\kappa_{2}^i}^2 - {\kappa_{\perp}^i}^2& -2\kappa_{2}^i s_2^i& 0 & 0\\
		2\kappa_{2}^i s_2^i &  {s_2^i}^2 -{\kappa_{2}^i}^2 - {\kappa_{\perp}^i}^2 & 0 & 0\\
		0 & 0 & {s_2^i}^2 -{\kappa_{2}^i}^2- {\kappa_{\perp}^i}^2& -2\kappa_{2}^i s_2^i\\
		0 & 0 & 2\kappa_{2}^i s_2^i &  {s_2^i}^2 -{\kappa_{2}^i}^2 - {\kappa_{\perp}^i}^2\\
	\end{bmatrix}.\notag\\
\end{align}
Once $\bm U^i$ is known, $\bm B^i$ is derived using Eq. (\ref{eq:linear_ns_adm}) such that
\begin{equation}
		\bm B^i = \Ha^{-2}{\begin{bmatrix} 
			{\mathbfitsf\Dz^i_1}^{-1} \left[\Rv\mathbfitsf\Dt - \mathbfitsf{M}_1^i \right] & 0\\
			0 &   {\mathbfitsf\Dz^i_2}^{-1} \left[\Rv\mathbfitsf\Dt - \mathbfitsf{M}_2^i \right] \\	
	\end{bmatrix}} 	\bm U^i .\label{eq:passage_U_vers_B}
\end{equation}
Note that the dependence of the solutions $\bm U^i$ and $\bm B^i$ on $\Rv$ and $\Rn$ is only implicit in equations \eqref{eq:BC_forcing} and \eqref{eq:passage_U_vers_B}, through the wavenumbers and attenuation coefficients $\left\{      \kappa_1^i,\,\kappa_2^i\right\}$ and $\left\{s_1^i,\,s_2^i\right\}$, obtained from the dispersion relation \eqref{eq:disp_relation} where they appear explicitly.}

\section{Expression of the coefficients $\left\{E_l^i,\mathit\Phi_l^i\right\}_{l=1..8}$}\label{App:exp_for_E_Phi}

{
Here we express explicitly the unknown coefficients $\left\{E_l^i,\mathit\Phi_l^i\right\}_{l=1..8}$ for the radial electric field $E_r$ and for the radial component of the potential gradient $\partial_r \phi$, respectively. As a reminder, $E_r$ and $\partial_r\phi$ take the form
\begin{align} \label{sol:gradPhi_E_Bessel_expension}
	\left\{E_r,\partial_r\phi\right\}\left(r,z,t\right)= \sum_{i=1}^{N_\perp}\left\{E_r^i,\partial_r\phi^i\right\}\left(z,t\right) J_1\left(\kappa_\perp^i r\right),
\end{align}
with 
\begin{align}
	\left\{E_r^i,\partial_r\phi^i\right\}\left(z,t\right)=& \left\lbrace\mathrm{e}^{s_1^i\,z} \left(\left\{E_1^i,\mathit\Phi_1^i\right\}\cos \left(t + \kappa_{1}^i\,z\right) - \left\{E_2^i,\mathit\Phi_2^i\right\} \sin\left(t + \kappa_{1}^i\,z\right) \right)\right.\notag\\
	&+  \mathrm{e}^{-s_1^i\,z} \left(\left\{E_3^i,\mathit\Phi_3^i\right\} \cos \left(t - \kappa_{1}^i\,z\right) - \left\{E_4^i,\mathit\Phi_4^i\right\}\sin\left(t - \kappa_{1}^i\,z\right) \right) \notag\\
	&+ \mathrm{e}^{s_2^i\,z} \left(\left\{E_5^i,\mathit\Phi_5^i\right\}\cos \left(t + \kappa_{2}^i\,z\right) - \left\{E_6^i,\mathit\Phi_6^i\right\} \sin\left(t + \kappa_{2}^i\,z\right) \right) \notag \\
	&+ \left. \mathrm{e}^{-s_2^i\,z} \left(\left\{E_7^i,\mathit\Phi_7^i\right\} \cos \left(t - \kappa_{2}^i\,z\right) - \left\{E_8^i,\mathit\Phi_8^i\right\} \sin\left(t - \kappa_{2}^i\,z\right) \right) \right\rbrace.\label{app_e}
\end{align}
so that we define the coefficient vectors $\bm E^i = \begin{bmatrix}  
	E_{l=1}^i &...& E_{l=8}^i
\end{bmatrix}^\mathrm{T}$ and $ \bm{\mathit{\Phi}}^i = \begin{bmatrix}  
	\mathit\Phi_{l=1}^i &...& \mathit\Phi_{l=8}^i
\end{bmatrix}^\mathrm{T}$.
Using the bock arrays (\ref{eq:block_array_space_derivative}-\ref{eq:block_array_Laplacian}) defined in appendix \ref{App:exp_for_U_B}, Eq. (\ref{eq:exp_Er}) and Eq. (\ref{eq:exp_radial_gradPhi}), we readily obtain the expression in term of $\bm U^i$ and $\bm B^i$ for $\bm E^i$,
\begin{equation}
	{\bm E^i} = -{\bm U^i} -\begin{bmatrix}
		\mathbfitsf\Dz^i_1 & 0 \\
		0 & \mathbfitsf\Dz^i_2\\
	\end{bmatrix} {\bm B^i}
\end{equation}
and $\bm{\mathit{\Phi}}^i$,
\begin{equation}
	{\bm{\mathit{\Phi}}^i} = -\begin{bmatrix}
		\mathbfitsf\Dz^i_1 - \mathbfitsf\Dt{\mathbfitsf{M}_1^i}^{-1}\mathbfitsf\Dz^i_1 & 0 \\
		0 &\mathbfitsf\Dz^i_2 - \mathbfitsf\Dt{\mathbfitsf{M}_2^i}^{-1}\mathbfitsf\Dz^i_2\\
	\end{bmatrix}  {\bm B^i} -{\bm U^i}.
\end{equation}
}

\bibliographystyle{jfm}
\bibliography{biblio}

\end{document}